\documentclass[twocolumn]{aastex62}

\usepackage{graphicx}
\usepackage{mathtools}
\usepackage{multirow}
\usepackage{bm}
\usepackage{gensymb}

\newcommand{\eppsilon}{$\varepsilon$ppsilon}

\def\UMelbourne{\affil{The University of Melbourne, School of Physics, Parkville, VIC 3010, Australia}}

\def\ASTRO{\affil{{ARC Centre of Excellence for All Sky Astrophysics in 3 Dimensions (ASTRO 3D)}}}

\def\UW{\affil{{University of Washington, Department of Physics, Seattle, WA 98195, USA}}}

\newcommand{\ASU}{\affil{{Arizona State University, School of Earth and Space Exploration, Tempe, AZ 85287, USA}}}

\def\UWeSci{\affil{{University of Washington, eScience Institute, Seattle, WA 98195, USA}}}

\def\Curtin{\affil{{International Centre for Radio Astronomy Research, Curtin University, Perth, WA 6845, Australia}}}

\def\UWAstro{\affil{{University of Washington, Department of Astronomy, Seattle, WA 98195, USA}}}

\def\Brown{\affil{{Brown University, Department of Physics, Providence, RI 02912, USA}}}

\def\Kumamoto{\affil{{Faculty of Science, Kumamoto University, 2-39-1 Kurokami, Kumamoto 860-8555, Japan}}}

\def\Dunlap{\affil{{Dunlap Institute for Astronomy and Astrophysics, University of Toronto, ON M5S 3H4, Canada}}}

\def\MIT{\affil{{MIT Kavli Institute for Astrophysics and Space Research, Cambridge, MA 02139, USA}}}

\def\CASS{\affil{{CSIRO Astronomy and Space Science (CASS), PO Box 76, Epping, NSW 1710, Australia}}}

\def\RRI{\affil{{Raman Research Institute, Bangalore 560080, India}}}

\graphicspath{{./}{images/}}

\received{January 1, 2018}
\revised{January 7, 2018}
\accepted{\today}
\submitjournal{ApJ}

\shorttitle{Improving MWA EoR Results}
\shortauthors{Barry et al.}

\begin{document}

\title{Improving the EoR Power Spectrum Results from MWA Season 1 Observations}

\author{N.~Barry}
\email{nichole.barry@unimelb.edu.au}
\UMelbourne
\ASTRO

\author{M.~Wilensky}
\UW

\author{C.~M.~Trott}
\Curtin
\ASTRO

\author{B.~Pindor}
\UMelbourne
\ASTRO

\author{A.~P.~Beardsley}
\ASU

\author{B.~J.~Hazelton}
\UW
\UWeSci

\author{I.~S.~Sullivan}
\UWAstro

\author{M.~F.~Morales} 
\UW

\author{J.~C.~Pober}
\Brown

\author{J.~Line}
\Curtin
\ASTRO

\author{B.~Greig}
\UMelbourne
\ASTRO

\author{R.~Byrne}
\UW

\author{A.~Lanman}
\Brown

\author{W.~Li}
\Brown

\author{C.~H.~Jordan}
\Curtin
\ASTRO

\author{R.~C.~Joseph}
\Curtin
\ASTRO

\author{B.~McKinley}
\Curtin
\ASTRO

\author{M.~Rahimi}
\UMelbourne
\ASTRO

\author{S.~Yoshiura}
\Kumamoto

\author{J.~D.~Bowman}
\ASU

\author{B.~M.~Gaensler}
\Dunlap

\author{J.~N.~Hewitt}
\MIT

\author{D.~C.~Jacobs}
\ASU

\author{D.~A.~Mitchell} 
\CASS
\ASTRO

\author{N.~Udaya~Shankar}
\RRI

\author{S.~K.~Sethi}
\RRI

\author{R.~Subrahmanyan}
\RRI
\ASTRO

\author{S.~J.~Tingay}
\Curtin
\ASTRO

\author{R.~L.~Webster}
\UMelbourne
\ASTRO

\author{J.~S.~B.~Wyithe}
\UMelbourne
\ASTRO



\begin{abstract}


Measurements of 21\,cm Epoch of Reionization structure are subject to systematics originating from both the analysis and the observation conditions. Using 2013 data from the Murchison Widefield Array, we show the importance of mitigating both sources of contamination. A direct comparison between results from \citet{beardsley_first_2016} and our updated analysis demonstrates new precision techniques, lowering analysis systematics by a factor of 2.8 in power. We then further lower systematics by excising observations contaminated by ultra-faint RFI, reducing by an additional factor of 3.8 in power for the zenith pointing. With this enhanced analysis precision and newly developed RFI mitigation, we calculate a noise-dominated upper limit on the EoR structure of $\Delta^2 \leq 3.9 \times 10^3$\,mK$^2$ at $k=0.20$\,\textit{h}\,Mpc$^{-1}$ and $z=7$ using 21\,hrs of data, improving previous MWA limits by almost an order of magnitude. 

\end{abstract}




\section{Introduction}
\label{sec:intro}

The evolution of structure in the early universe holds significant value to our understanding of cosmology and astrophysics. However, it remains unobserved. Of particular interest is the Epoch of Reionization (EoR), the time period where near-uniform neutral hydrogen coalesced into stars and galaxies. Observations of extragalactic sources (e.g \citealt{robertson_early_2010}) and the cosmic microwave background (CMB; e.g. \citealt{adam_planck_2016}) have put some constraints on the neutral hydrogen fraction, though not much is known beyond an approximate timing of when the EoR occurred.

A direct detection of the structure of neutral hydrogen, and how it evolves in time, is possible using the 21\,cm hyperfine transition. The measured frequency corresponds to a line-of-sight distance due to the narrow emission width, and thus the structure of the EoR can be measured as a function of time. Many collaborations are seeking an EoR 21\,cm detection, including the Giant Metrewave Radio Telescope \citep{paciga_gmrt_2011}, LOw Frequency Array (LOFAR; \citealt{yatawatta_initial_2013, van_haarlem_lofar:_2013}), Precision Array for Probing the Epoch of Reionization (PAPER; \citealt{parsons_precision_2010}), Hydrogen Epoch of Reionization Array \citep{pober_what_2014, deboer_hydrogen_2017}, and the Murchison Widefield Array (MWA; \citealt{bowman_science_2013,tingay_murchison_2013,wayth_phase_2018}).

A recent sky-averaged detection of the early EoR using the Experiment to Detect the Global EoR Signature, (EDGES; \citealt{bowman_absorption_2018}) revealed an absorption profile twice as bright as predicted, thereby suggesting new physics. While this detection does not probe structure, it sets a precedent for the potential impact of direct measurements of the EoR. So far, there have been no detections of the EoR structure due to the challenges associated with analysis and spectral accuracy.

Astrophysical foregrounds are many orders of magnitude brighter than the EoR signal. Therefore, our ability to measure EoR structure depends on how well we can separate the foregrounds from the signal in analysis. Power spectrum space naturally separates spectrally smooth foregrounds from the predicted spectral variation of the EoR signal while quickly averaging down the thermal noise.

Recent discoveries of analysis signal loss have changed the state of the field and reemphasized the importance on analysis pipeline efficacy and signal loss simulations \citep{cheng_characterizing_2018}. As a result, the best published upper limits on EoR structure have remained close to ${\small \sim} 10^4$\,mK$^2$ \citep{paciga_simulation-calibrated_2013,dillon_empirical_2015,beardsley_first_2016,patil_upper_2017}, with the most recent competitive MWA limits as $\Delta^2 \leq 2.7 \times 10^4$\,mK$^2$ at $k = 0.27$\,\textit{h}\,Mpc$^{-1}$ and $z = 7.1$ published by \citet{beardsley_first_2016}.

In this work, we first improve on the best MWA limits through pure analysis techniques alone. We begin with the identical 32\,hr data set and the same code packages as \citet{beardsley_first_2016}, with the only difference between the analyses being our new precision techniques. This highlights the importance of our analysis pipeline, proving that the dominant systematic preventing lower limits was the spectral accuracy of our techniques. Then, we further improve upon this limit through additional systematic mitigation using the latest RFI-detection methodologies, reducing the data set to 21\,hr.

We achieve a new EoR upper limit at $k=0.20$\,\textit{h}\,Mpc$^{-1}$ and $z=7$ of $\Delta^2 \leq 3.9 \times 10^3$\,mK$^2$. This limit is systematic dominated, and would not benefit significantly from further integration. However, the zenith-pointing subset gives a similar limit, which \textit{is} noise dominated, indicating pointing or beam-related errors as a dominant systematic in our analysis. We do not expect to be able to detect the EoR with less than a few hundred hours of integration \citep{beardsley_eor_2013}, but a smaller integration is still capable of proving the viability of our analysis and foreground mitigation techniques.

Descriptions of the MWA instrument and data sets used in this work are given in \S\ref{sec:MWA_obs}. Then, we briefly summarize the major components of the Fast Holographic Deconvolution (FHD)/Error Propagated Power Spectrum with Interleaved Observed Noise ({\eppsilon}) data analysis pipeline in \S\ref{sec:pipeline}, focusing on recent improvements. In \S\ref{sec:AB2016}, we compare our updated analysis to \citet{beardsley_first_2016} by reducing the same data set. We further improve this limit in \S\ref{sec:new_limit} by using the latest RFI-mitigation techniques to select a subset of the data, and we provide validation of this new limit in \S\ref{sec:checks}. Finally, we discuss potential future improvements in \S\ref{sec:future}.

\section{Observations and the MWA}
\label{sec:MWA_obs}
To understand the most recent improvements to the EoR upper limit, we must first briefly describe the main instrumental aspects of the MWA. We also summarize the data preprocessing, focusing on our nominal RFI flagging, as well as detail the data sets used in this work.

\subsection{The MWA}
\label{subsec:MWA}

The MWA is a radio interferometer located on a designated radio-quiet site in Western Australia. It has a dense cluster of elements at the core, which allows for extra sensitivity on EoR modes, and it has a pseudo-random scatter of elements at longer baselines for imaging. Due to its design, there is an opportunity to utilize MWA-made catalogs for instrumental calibration and foreground subtraction in EoR analysis. Upgrades to the MWA have separated the scientific cases into two independent layouts \citep{wayth_phase_2018}, however, all data reduced in this paper uses the original Phase I layout \citep{tingay_murchison_2013}.

The sky voltage is observed with 128 elements of 16 dual-polarization dipoles in a 4$\times$4 layout over a ground screen. Analog delays are added to the signal path to point each dipole toward the target, and then each set of 16 dipoles is beamformed to form one element. The signal then travels to the receiver, where it is digitized in a first-stage coarse frequency channelizer. This introduces aliasing every 1.28\,MHz which must be removed via flagging in the analysis pipeline \citep{prabu_digital-receiver_2015}. Each 1.28\,MHz channel is multiplied by a quantized gain to achieve a flatter response and to avoid bit quantization errors. A bandwidth of 30.72\,MHz is then selected and transported to the correlator, where a second-stage channelizer creates 10\,kHz channels. The signals are then correlated and averaged to a specified time and frequency set, specifically 0.5\,s and 40\,kHz for a 112\,s interval per observation in this work.

We observe a few targeted fields of the sky for EoR science. This is achieved through a ``drift and shift'' strategy, where the instrument is pointed toward the target field every 30 minutes using analog delays \citep{trott_comparison_2014}. In between shifts, the pointing remains constant and each new pointing results in a different beam shape and sampling of the sky. For EoR science, we only reduce data from pointings near zenith to minimize the effect of beam modeling errors. We use two pointings before zenith to two pointings after zenith in this work, targeting the ``EoR0'' field at R.A. $= 0.00^\mathrm{h}$, decl. $= -27\degree$. This field is naturally low in sky temperature due to minimal foreground emission, and thus a great candidate for EoR observing.

\subsection{Preprocessing}
\label{subsec:preprocess}

We run a pre-pipeline package to generate input data files that are RFI-flagged and frequency/time averaged. This simultaneously reduces data size and mitigates adverse RFI effects. We use \textsc{cotter}, which subsequently calls the flagging package \textsc{aoflagger}\footnote{\url{https://sourceforge.net/p/aoflagger/wiki/Home/}} \citep{offringa_low-frequency_2015}.

We input the 0.5\,s and 40\,kHz sampled data into \textsc{aoflagger}, which performs RFI flagging at this highest resolution. Line-like RFI features are found via a summed frequency and time threshold method after the sky contribution is estimated in the visibilities \citep{offringa_post-correlation_2010}. Then, a time and frequency morphological technique is used to catch broadband and variable features \citep{offringa_morphological_2012}.

We perform further flagging based off of known MWA instrumental effects. In particular, we flag the first two seconds and the last four seconds of an observation to mitigate beamformer lag and potential for dropped integration blocks in some 2013 data, respectively. For frequency, we flag the first two and last two frequency channels per coarse band to avoid channelizer aliasing, as well as the center DC channel per coarse band to avoid adverse bit rounding errors.

Averaging is then performed to output 80\,kHz frequency and 2\,s time resolution in a standard UVFITS format.

\subsection{Data sets}
\label{subsec:data_set}

A 32\,hr data set of observations was used to calculate EoR limits for the MWA in \citet{beardsley_first_2016}, and we will use the same data set to verify our pipeline improvements in \S\ref{sec:AB2016}. This data set measures approximately 167--197\,MHz from 2013 August until November. The target field is EoR0, and only pointings near zenith are included.

These 1029 observations were originally chosen for their clean statistics; they passed multiple mitigation tests, including pointing-based cuts, window power ratio cuts, polarization difference cuts, and residual image RMS cuts (see \citealt{beardsley_first_2016} for quality control techniques). They represent the best data at the time for MWA EoR analysis.

In addition to our pipeline-verification data sets, we apply improved quality assurance techniques to remove additional contaminated data from the original 1029 observations to generate our best EoR upper limits in \S\ref{sec:new_limit}. All data with signatures of digital TV, amounting to 311 observations, are completely removed. A further 40 observations are removed by cutting data with RFI occupancy fraction greater than 40\%. This creates a subset of about 21\,hr, or 678 observations, from the previous data set. A description of our improved RFI-detection techniques is provided in \S\ref{subsec:data_select}.

As another part of our pipeline verification, we compare a small data set analyzed with FHD/{\eppsilon} to an analysis from RTS/CHIPS \citep{mitchell_real-time_2008,trott_chips:_2016} in \S\ref{subsec:pipe_compare}. This small data set only consists of the zenith pointing of the 678-observation data set, amounting to approximately 5\,hr.

\section{The FHD/$\mathcal{E}$ppsilon pipeline}
\label{sec:pipeline}

Accuracy in analysis is a crucial aspect to EoR power spectrum measurements. Every element of our data analysis, including calibration, foreground subtraction, observation integration, and power spectrum estimation is highly bespoke to the accuracy requirements necessary to detect the EoR.

There are two separate, open-source packages in our pipeline: FHD\footnote{\url{https://github.com/EoRImaging/FHD}} (Fast Holographic Deconvolution) and {\eppsilon}\footnote{\url{https://github.com/EoRImaging/eppsilon}} (Error Propagated Power Spectrum with Interleaved Observed Noise). FHD is an implementation of an efficient deconvolution algorithm \citep{sullivan_fast_2012}, though now we use its precision analysis capabilities to calibrate and image each observation without deconvolution \citep{barry_fhd/eppsilon_2019}. {\eppsilon} calculates the resulting power spectrum and noise estimates from an integration of many observations. This general setup shares similar aspects with other imaging pipelines \citep{paciga_gmrt_2011,patil_constraining_2014,shaw_all-sky_2014,shaw_coaxing_2015,dillon_empirical_2015,ewall-wice_first_2016,jacobs_murchison_2016,trott_chips:_2016,patil_upper_2017}. However, a key focus of FHD/{\eppsilon} is end-to-end error propagation for error estimation. This is relatively uncommon amongst power spectrum estimators, but it is an integral feature in both {\eppsilon} and CHIPS \citep{trott_chips:_2016}.

In this section, we summarize the main computational tasks and outputs of the FHD/{\eppsilon} pipeline, highlighting the significant accuracy improvements we have made since \citet{beardsley_first_2016} and \citet{jacobs_murchison_2016}. Specifically, our pipeline is an image-based analysis that generates a reconstructed power spectrum \citep{morales_understanding_2019}. The resulting errors we aim to reduce are thus specific to this style of analysis. 

\begin{figure}
\centering
	\includegraphics[width=.75\columnwidth]{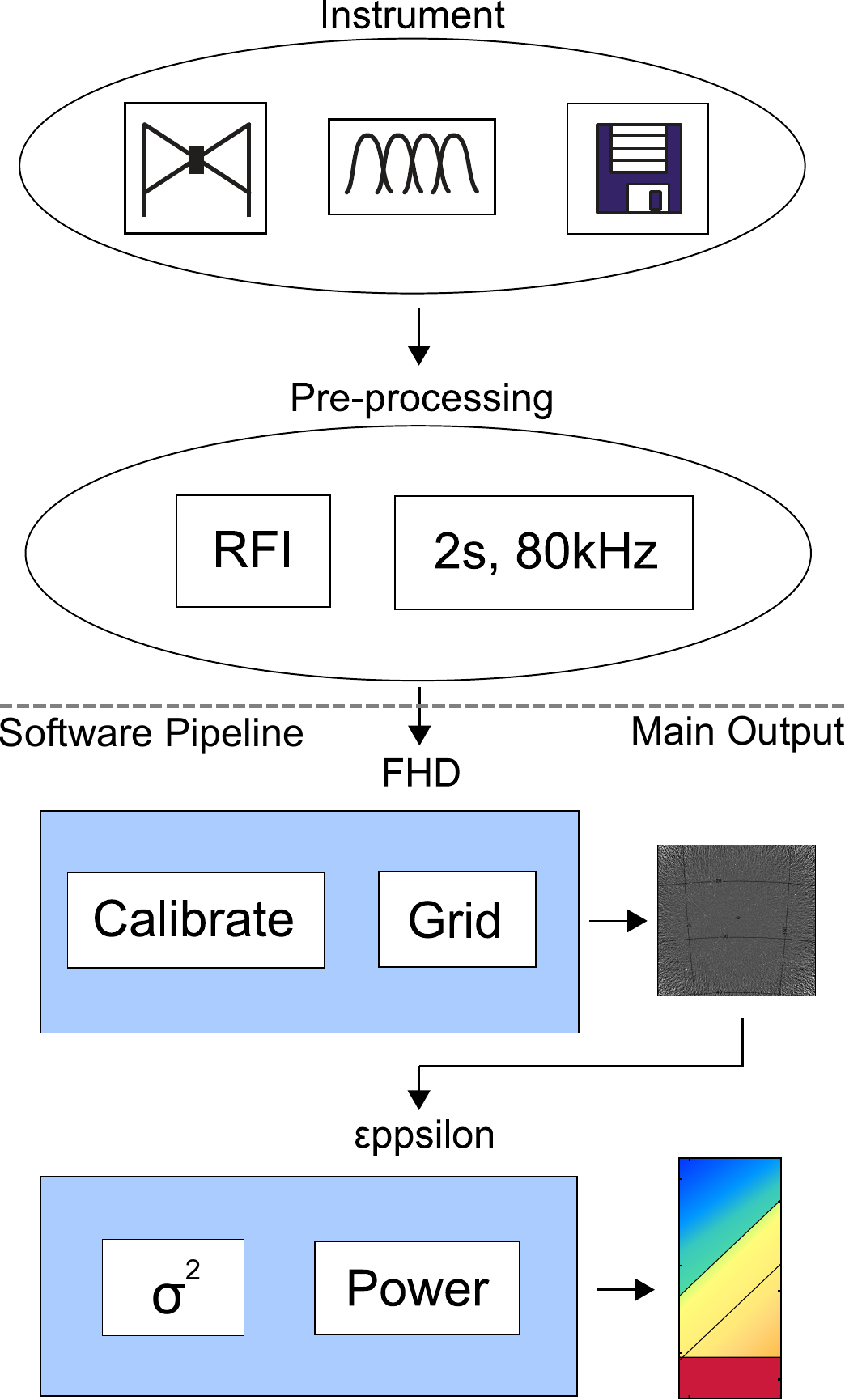}
	\caption{The data flow from measurement to power spectrum. There are four main blocks: (1) measurement, electronics, and raw data output from the instrument (\S\ref{subsec:MWA}), (2) RFI detection and averaging during preprocessing (\S\ref{subsec:preprocess}), (3) calibration, gridding, and image creation during the FHD analysis (\S\ref{subsec:FHD}), and (4) error propagation, power spectrum estimation, and binning during the {\eppsilon} analysis (\S\ref{subsec:eppsilon}).}
	\label{fig:flow}
\end{figure}

Figure~\ref{fig:flow} provides context for how our analysis packages fit into the data flow. For a full description of the tasks and outputs of the FHD/{\eppsilon} pipeline, please see \citet{barry_fhd/eppsilon_2019}.

\subsection{FHD}
\label{subsec:FHD}

In brief, FHD calculates calibrated images from measured visibility data. Various transformations and assumptions must take place to achieve these results, therefore the narrative of our data reduction is that of accuracy and precision. We summarize the main steps in the analysis, focusing on recent improvements.

First, we estimate the dipole response using an image-space beam from simulated Jones matrices for each frequency. Then, we build the response of the beamformed element and transform it into the $\{u,v,f\}$-domain, or the space of the visibility measurements. However, there is an accuracy limitation to simulations which model the Jones matrices at low dB. To account for this, we build a mask in $uv$-space to cut at this low dB level, and we renormalize the beam such that the edge smoothly goes to zero. This recent improvement ensures a smoother beam in Fourier space, and reduces systematics in sensitive EoR modes.

This beam kernel, or $uv$-beam response, is as instrumentally accurate and smooth as possible, and thus will be used in calibration. However, we now also use a modified gridding kernel in power spectrum estimation, analogous to the concept of a Tapered Gridded Estimator \citep{choudhuri_visibility-based_2014, choudhuri_tapering_2016}. We apply the square of a Blackman-Harris window to the beam image. This acts as an image-space weighting to the observation when normalizations are properly taken into account. However, this creates correlations between image pixels; we investigate the importance of this effect in the power spectrum in \S\ref{subsec:uncertainty}. In this work, we refer to the kernel matched to the instrument beam as the ``instrumental gridding kernel" and the kernel incorporating the image domain window function as the ``modified gridding kernel." 

Next, we build model visibilities for calibration. We generate a model $uv$-plane at half-wavelength resolution by performing a discrete Fourier Transform (DFT) of the flux and position of over 10,000 sources to a specified set of grid points in $uv$-space. At every baseline location in $uv$-space, we multiply the model $uv$-plane by our instrumental gridding kernel and integrate the result. This generates model visibilities, or an estimate of what the instrument measured.

We are reaching levels of precision that are affected by finite sum integration errors, especially in the process of estimating model visibilities \citep{kerrigan_improved_2018}. Since the modified gridding kernel is a convolution in $uv$-space, it smooths out these errors as a function of frequency. This has important consequences in power spectrum space, and thus we use the modified gridding kernel when making power spectra and the instrumental gridding kernel when calibrating. This results in two sets of model visibilities: one for calibration that is instrumentally accurate, and one for power spectra that is frequency-smooth.

The introduction of the GaLactic and Extragalactic All-sky Murchison Widefield Array (GLEAM) survey in recent years has greatly increased the number of point sources in our sky model \citep{hurley-walker_GLEAM}. As a result, this has improved the accuracy of our sky-based calibration and the dynamic range in the power spectrum. Since our absolute flux scale has changed with GLEAM, our absolute power spectrum normalization has changed as well. Future works that scientifically compare EoR upper limits will need to match absolute flux scales and incorporate flux scale uncertainties.

We then calibrate the measured visibilities using the model visibilities in a sky-based calibration. An alternating direction implicit method is used to minimize a $\chi$-squared equation to determine the gains from each element, which are assumed to account for the difference between the data and model \citep{mitchell_real-time_2008,salvini_fast_2014}. This necessitates several assumptions, including independence in time, elements, polarization, and frequency.

These assumptions have consequences in the power spectrum. In \citet{beardsley_first_2016}, we employed a method to smooth the bandpass gains as a function of frequency to reduce spectral structure. In our current implementation, we use the auto-visibilities to calculate the resulting amplitude per element. The auto-visibilities do not measure any structure on the sky, and thus are not subject to the incomplete sky errors as a function of frequency in \citet{barry_calibration_2016}. However, the noise statistics and the overall amplitude will not be the same as the cross-visibilities. We continue to use the fit cross-phases with a cable reflection in calibration, as the auto-visibilities do not contain any phase information. For a full characterization of our calibration, please see \citet{barry_fhd/eppsilon_2019}.

Errors in the bit statistics can affect both the cross- and auto-visibilities if there is bit over or undersaturation \citep{barry_thesis}. We have found that in 2013 data, the upper part of the band (${\small \sim}$187.5 -- 197\,MHz) was adversely affected by the bit statistics. An improperly tuned channel gain resulted in bit oversaturation, leading to an uncorrectable bias in the data. We now completely flag this part of the band in 2013 data, which limits the usable redshifts in power spectrum measurements.

Once we have calibrated, we grid the data and model visibilities onto the $uv$-plane. Each visibility value per frequency is multiplied by our modified gridding kernel and pixelated onto a regular $uv$-grid. In addition, we also grid beam kernels of integrated value 1 and squared beam kernels of integrated value 1 to separate $uv$-planes. These two additional $uv$-planes are the sampling map and variance map, respectively, and will be used in {\eppsilon} for power spectrum estimation and error propagation. In order to generate noise estimates in {\eppsilon}, we split each $uv$-plane into two interleaved time steps.

Since we did not incorporate $w$-projection effects \citep{cornwell_noncoplanar_2008}, our gridded $uv$-planes are not in a basis that is coherent across observations. Therefore, we choose to Fourier transform the sampling map, the variance map, and the unweighted data $uv$-planes separately into image space. We interpolate to HEALPix projection,\footnote{HEALPix: the Hierarchical Equal Area isoLatitude Pixelization of a sphere \citep{gorski_healpix:_2005}.} which is a fixed basis and therefore integrable across observations. 

When we use image space to integrate, we are subject to aliasing effects due to a limited extent in the image. Our modified gridding kernel acts as a window filter, which greatly reduces image aliasing effects. In order to retain similar effective area to our previous 560\,deg$^2$, we now image 8090\,deg$^2$.

\subsection{{\eppsilon}}
\label{subsec:eppsilon}

Our power spectrum pipeline, {\eppsilon}, calculates the data power, model power, residual power, observed noise, expected noise, and uncertainty estimates from integrated images. We determine these various data and noise products to build reliable upper limits.

First, {\eppsilon} transforms each of the integrated images back into the $\{u,v,f\}$-domain. This includes the separate interleaved time samples for the data, the sampling map, and the variance map for each polarization. Each data product is then weighted with the sampling map, and similarly the variance map is weighted with the square of the sampling map.

At this point, we choose to perform weighted sums and differences of the data to build both a propagated noise spectrum and a power spectrum simultaneously. Using our interleaved time samples, we calculate maximum likelihood estimates of the mean and the noise by using the sampling-map-weighted variances as the new weights. This weighting scheme is also used to calculate maximum-likelihood uncertainty estimates. These uncertainty estimates are compared to $T_{\textrm{sys}}$ calculations propagated directly from the time-interleaved visibility differences, and thus we have consistent uncertainty estimates for every quantity within the full FHD/{\eppsilon} pipeline. 

To continue onto power spectra, we must transform into the $\{k_x,k_y,k_z\}$-domain. While $u$ and $v$ are easily converted into $k_x$ and $k_y$ via cosmological parameters, the transform from $f$ into $k_z$ is nuanced. Our frequency sampling is irregular: baselines move as a function of frequency and we have flagged frequencies due to RFI and instrumental systematics. Therefore, we perform a Lomb--Scargle periodogram \citep{lomb_least-squares_1976,scargle_studies_1982} to find an orthogonal basis. The phase is not preserved in the Lomb--Scargle, thus restricting its use to power spectrum calculations and destroying information useful to the bispectrum \citep{bharadwaj_probing_2005}, which may prove to be a vital statistic in the future \citep{majumdar_quantifying_2018,trott_gridded_2019,watkinson_21-cm_2019}.

Immediately before we perform the frequency transform, we first remove the mean of the amplitude. This reduces contamination from the bright, intrinsic foregrounds coupling into higher $k_z$-modes during the transform. We add this term back in as the DC component in the $\{k_x,k_y,k_z\}$-domain to preserve power. This new average-removal method improves the distribution of power in 2D and 1D power spectrum compared to previous analyses.

We now construct our cross power estimation from the maximum-likelihood mean and noise of the 3D $\{k_x,k_y,k_z\}$-cube. The power of the mean minus the power of the noise, divided by four, gives the same power estimation as the cross-power between the interleaved time samples \citep{barry_fhd/eppsilon_2019}.

Since we choose to form the cross power after transformations, we are able to easily propagate our noise throughout {\eppsilon}. We assume negligible cross correlation between pixels, hence the propagation is simple sum/difference error propagation. We compare this propagated expected noise to the observed noise calculated from even--odd differences in \S\ref{subsec:uncertainty} to test this assumption. This is particularly important since increasing the integrated image area increases the $uv$-resolution.

In order to calculate upper limits and related products from the 3D $\{k_x,k_y,k_z\}$-cube, we calculate maximum-likelihood weighted averages of the bins. To form 2D power spectra as a function of $k$-modes perpendicular to the line of sight, $k_\perp$, and $k$-modes parallel to the line of sight, $k_{||}$, we average in cylindrical regions. To form 1D power spectra as a function of $|k|$, we average in spherical regions.

The final outputs of our FHD/{\eppsilon} pipeline are 2D and 1D power spectra products of the calibrated data, model data, residual data, expected noise, observed noise, and error bars. These various products will be used to determine our upper limits in \S\ref{sec:AB2016} and \S\ref{sec:new_limit} and to provide evidence of signal preservation and proper error propagation in \S\ref{sec:checks}.

\section{Direct limit comparison}
\label{sec:AB2016}

In \citet{beardsley_first_2016}, the FHD/{\eppsilon} pipeline was used to reduce 32\,hr of MWA data for an EoR upper limit. The data reduction was dominated by systematics, and thus larger integrations would not have benefited the analysis significantly. At the time, it was unknown whether the limiting systematics were related to the data or to the analysis.

Many improvements have been made to the FHD/{\eppsilon} pipeline since then, as outlined in \S\ref{sec:pipeline}. By reducing the same data set, we can determine how significant these improvements are toward lowering the MWA EoR limit. This will also help identify the nature of the dominating systematics in \citet{beardsley_first_2016}.

Our updated analysis completely flags the upper part of the band due to an improperly tuned channel gain. To make a direct comparison, we rebin the data in \citet{beardsley_first_2016} and in our updated analysis to include the range 168.555--187.275\,MHz. We avoid frequency regions where the bit statistics indicate truncation or saturation, and are thus different between the auto-visibilities and the cross-visibilities. This includes the first coarse band (167.115--168.235\,MHz) and the upper part of the band (187.595--197.675\,MHz). In addition, we avoid one extra channel each near these ranges (168.395 and 187.435\,MHz) since they have flagged contributions, which can have consequences in power spectrum space \citep{offringa_impact_2019}. With the Blackman-Harris window applied, we have an effective bandwidth of 9.4\,MHz at approximately redshift 7.

The calibration catalog, and thus the absolute flux scale, has changed between the two analyses. The previous catalog, KGS\footnote{KATALOGSS, the KDD (Knowledge Discovery in Databases) Astrometry, Trueness, and Apparent Luminosity of Galaxies in Snapshot Surveys, abbreviated as KGS.} \citep{carroll_KGS}, was matched to the flux density of MWA Commissioning Survey (MWACS; \citealt{hurley-walker_murchison_2014}). Their absolute flux scale is significantly higher due to residual flux and primary beam errors compared to our current catalog, GLEAM \citep{hurley-walker_GLEAM}. In order to have a fair comparison, we need to apply a scale factor to the \citet{beardsley_first_2016} analysis. We compare the mean calibration amplitudes for the frequency range of interest, and find that the previous analysis was 28\% brighter in power for the E--W polarization and 23\% brighter in power for the N--S polarization. We scale down both the \citet{beardsley_first_2016} upper limits and thermal noise by this correction factor for our comparison.

In \citet{beardsley_first_2016}, the integrated HEALPix image was limited to a square $\sim$20$\degree$ across, both to select the clean center of the antenna beam and to reduce computational costs. This hard image cut was found to cause aliasing in the $uv$-plane, and was replaced by the modified gridding kernel discussed in \S\ref{subsec:FHD}. This modified kernel performs the same task of selecting data from the center of the antenna beam, without the hard edge, but requires using an integrated HEALPix map that is over 10$\times$ larger in area. This change in integration scheme results in different levels of contamination in the thermal noise, simulated in \S\ref{subsec:uncertainty}. Therefore, our noise is about a factor of 3 lower than \citet{beardsley_first_2016} for the same data set.

\subsection{2D power spectrum comparison}
\label{subsec:2D_compare}

\begin{figure*}[h!]
\centering
	\includegraphics[width = \textwidth]{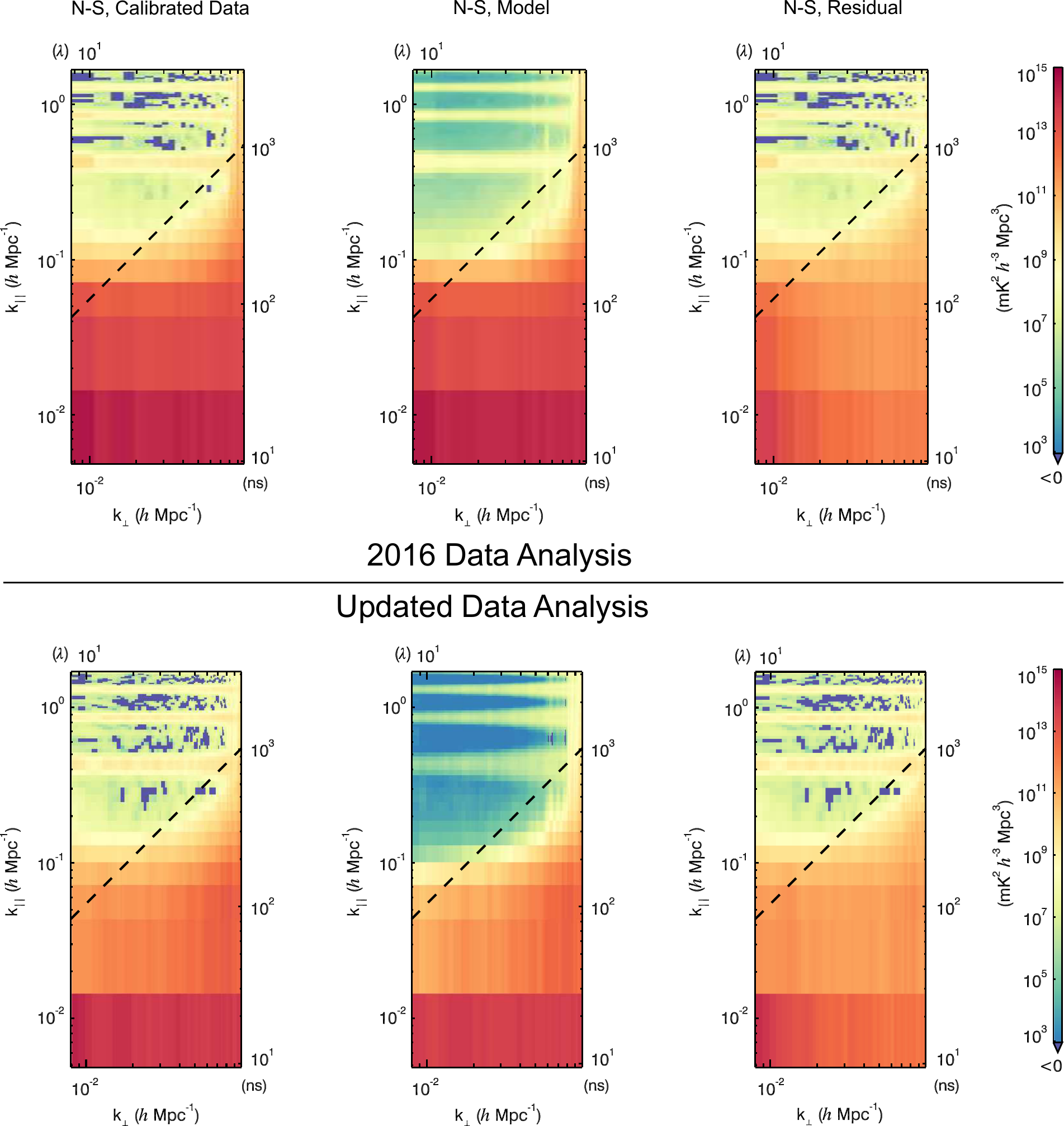}
	\caption{The 2D power spectra comparison between the \citet{beardsley_first_2016} analysis (top row) and our updated analysis (bottom row) with the same observation data set and binning scheme. The calibrated data (left column), the subtraction model (middle column), and the residual (right column) 2D power spectra are shown for the N--S polarization. Our updated precision techniques and improved calibration reduce foreground coupling into the EoR window (above the dashed line).}
	\label{fig:2d_AB_compare}
\end{figure*}

The 2D power spectrum is a useful diagnostic due to the characteristic contamination regions, and thus we can draw useful conclusions via their comparison. We create 2D power spectra for three data products: the calibrated data, the subtraction model, and the residual data after foreground removal.

We apply the frequency mask to both data sets, and then perform cylindrical averaging in $k_x$ and $k_y$ to generate $k_\perp$, or $k$-modes perpendicular to the line of sight. In addition, the $k_z$-component is relabeled $k_{||}$, or $k$-modes parallel to the line of sight.

Power contamination in the \{$k_\perp,k_{||}$\}-space occurs in distinctive regions. Foreground contamination is present at low $k_{||}$ since it does not vary quickly as a function of frequency. Instrument chromaticity also couples the foregrounds into higher $k_{||}$-modes along a constant slope called the ``foreground wedge'' \citep{datta_bright_2010,morales_four_2012,parsons_per-baseline_2012,trott_impact_2012,vedantham_imaging_2012,hazelton_fundamental_2013,pober_opening_2013,thyagarajan_study_2013,liu_epoch_2014}. Above this region is the ``EoR window,'' and this is where we expect to be able to make our measurements. Horizontal contamination lines constant in $k_{||}$ are caused from regular flagging of channelizer aliasing (described in \S\ref{sec:MWA_obs}).  For a full description of the 2D power spectrum from {\eppsilon}, please see \citet{barry_fhd/eppsilon_2019}.

Figure~\ref{fig:2d_AB_compare} shows the diagnostic 2D power spectra of the calibrated data, the model, and the residual for the N--S polarization for both the \citet{beardsley_first_2016} data reduction and our updated data analysis. There are several key differences.
\begin{enumerate}
  \item The power in the foreground wedge is less contained in the 2016 analysis; spectrally smooth foregrounds should contaminate the lowest $k_{||}$-mode by orders of magnitude more than other modes. This is an indication that foreground power was coupled to higher $k_{||}$-modes more than expected. We mitigate this in our updated analysis with improved calibration and the average-removal technique in power spectrum estimation.
  \item The model power in the updated analysis is much lower in the EoR window. This is a strong indication that our analysis is more precise, and couples less foreground power into the sensitive measurement modes than before. Many improved techniques described in \S\ref{sec:pipeline} contribute to this, including the modified gridding kernel.
  \item Contamination in the EoR window for both the calibrated data and residual is reduced in the updated analysis, especially between the horizontal flagging harmonics. This is most likely due to higher precision in the model and updated calibration techniques using auto-visibilities.
\end{enumerate}
The 2D comparison in Figure~\ref{fig:2d_AB_compare} demonstrates significant improvement in sensitive measurement regions and in the expected distribution of foreground power. Considerable advances have been made for the calibrated data and the model, resulting in an improved residual 2D power spectrum.

\subsection{1D upper limit comparison}
\label{subsec:1D_compare}

\begin{figure*}
\centering
	\includegraphics[width = \textwidth]{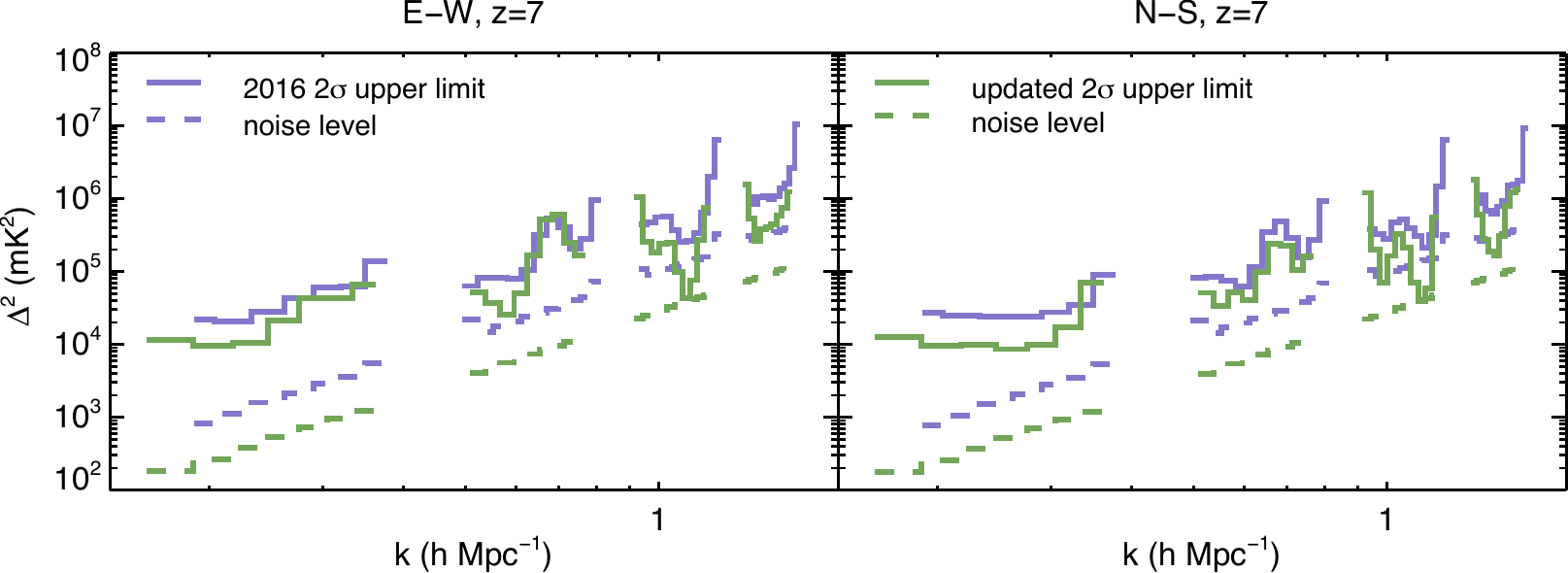}
	\caption{The 1D EoR upper limit comparison between \citet{beardsley_first_2016} (purple) and our updated analysis (green) for the E--W and N--S polarizations at a band centered on redshift 7 for the same 1029-observation data set. The dashed lines are the thermal noise levels of each analysis. Our updated analysis has less power contamination on most $k$-modes.}
	\label{fig:AB_limit_compare}
\end{figure*}

In order to show quantitative improvement, we calculate the 1D EoR upper limit between the two analyses. We use a similar mask and binning scheme to that presented in \citet{beardsley_first_2016} to have a direct comparison of the contamination levels in sensitive regions.

First, we exclude wavelengths in $k_\perp$ which are (1)~greater than 70\,$\lambda$ due to low $uv$-coverage and (2)~less than 10\,$\lambda$ due to increased contamination from coarse band harmonics. Then, we apply a $k_{||}$-mask to all bins lower than 0.15\,\textit{h}\,Mpc$^{-1}$ to avoid a systematic floor from foreground leakage. Finally, we remove the foreground wedge with a small buffer; we exclude along the horizon line with an increased slope of 14\% \citep{dillon_empirical_2015}. We also exclude $k_\perp$-bins associated with the coarse band harmonics, however, our analysis cannot exactly match \citet{beardsley_first_2016} due to a difference in resolution.

Figure~\ref{fig:AB_limit_compare} shows the comparison of the 1D EoR upper limits between the \citet{beardsley_first_2016} analysis and our updated analysis for the E--W and N--S polarizations at redshift 7. The limits are described as $\Delta^2(k) = k^3 P_{21}(k)/(2\pi^2)$ in units of mK$^2$. Again, the only difference between the two approaches is the analysis; the raw data and the $k$-space binning scheme have remained the same. The updated analysis from this work has a lower EoR upper limit for most $k$-modes.

In particular, the most sensitive measurement modes at small $k$ have lower EoR upper limits. The best mode in the N--S polarization from the rebinning of the \citet{beardsley_first_2016} analysis is $\Delta^2 \leq 2.37 \times 10^4$\,mK$^2$ at $k=0.25$\,\textit{h}\,Mpc$^{-1}$, while the best mode in the updated analysis is $\Delta^2 \leq 8.59 \times 10^3$\,mK$^2$. We have improved by a factor of $2.8$. Likewise, for the best modes in the E--W polarization, we have improved by a factor of $2.1$.

This is an indication that a major systematic in the \citet{beardsley_first_2016} analysis is related to the precision of the data reduction. There is a consistent improvement for most $k$-modes, suggesting that a contamination floor in the EoR window caused by foreground-coupled power has been mitigated by our new techniques. Figure~\ref{fig:2d_AB_compare} supports this conclusion; major power reduction in the model 2D power spectra, especially in the EoR window, indicates a significant improvement in analysis precision.

By analyzing the same data set in \citet{beardsley_first_2016} with our updated pipeline, we directly compared how our new precision techniques affected the 2D and the 1D power spectrum. Now, we have a consistent narrative that the FHD/{\eppsilon} pipeline has improved significantly, and that analysis systematics have been greatly reduced.

\section{Updated limit}
\label{sec:new_limit}

Our improvements to the FHD/{\eppsilon} pipeline (\S\ref{sec:pipeline}) lowered the previous MWA EoR upper limits significantly (\S\ref{sec:AB2016}). However, we can further lower these limits using new techniques to remove faint-RFI-contaminated observations. To calculate the best possible limits from 2013 data, we will choose a subset of the \citet{beardsley_first_2016} data set along with a new binning scheme. Using these 678 observations, we report an updated MWA EoR upper limit.

\subsection{Data selection}
\label{subsec:data_select}

We use \textsc{aoflagger} for primary RFI flagging in the FHD/{\eppsilon} pipeline (\S\ref{subsec:preprocess}). However, we are able to identify a substantial number of observations with leftover ultra-faint RFI contamination using \textsc{ssins}\footnote{https://github.com/mwilensky768/SSINS} \citep{wilensky_absolving_2019}. 

\textsc{ssins} operates by  time-differencing visibilities to subtract out the slowly varying sky, and then averaging the amplitudes of these visibility differences over the set of baselines in the array. This leaves a single dynamic spectrum per polarization in which flagging can be performed. This is called the sky-subtracted incoherent noise spectrum (SSINS) of the observation. At the expense of more finely grained baseline-to-baseline information, this gives a dramatic sensitivity boost that allows identification of RFI well below the thermal noise levels of a single baseline. In addition, a match-shape filter is implemented within the \textsc{ssins} framework to further boost sensitivity to known RFI contaminants such as digital television (DTV). 

While \textsc{ssins} can be used to flag time/frequency regions, we opted instead only to use it to catalog particular types and levels of contamination. This information was then used to cut out entire 112\,s observations with residual faint RFI, rather than try to recover the observations. Two main RFI cuts were made: (1) any observations containing DTV signals identified by the \textsc{ssins} match filter were removed, and (2) any observations where over 40\% of the SSINS samples\footnote{ Given instrumental effects in the MWA, we did not include coarse band edges in this classification.} were identified as contaminated by the \textsc{ssins} flagging procedures were removed. These procedures identified 311 observations with DTV and a further 40 observations with high levels of RFI occupancy. 

\begin{figure}[h!]
\centering
	\includegraphics[width = .7\columnwidth]{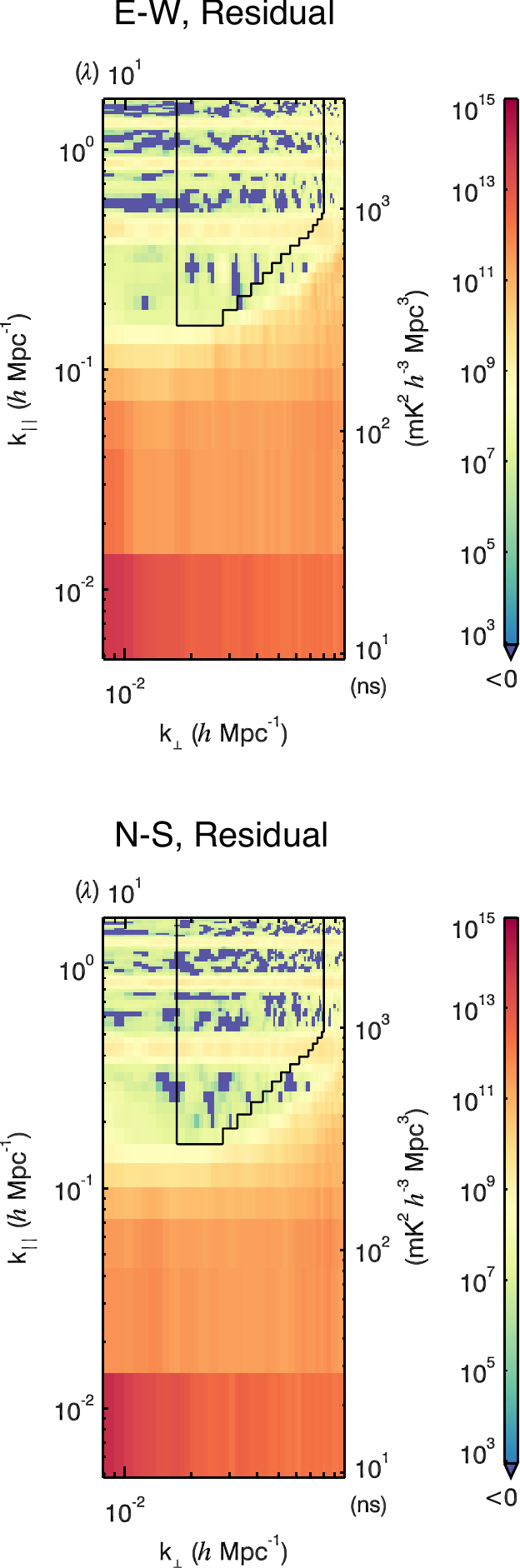}
	\caption{Residual 2D power spectra for the E--W and N--S polarizations for 678 observations selected with \textsc{ssins}. Contours show the region where we make $\{k_{||},k_{\perp}\}$ selections to avoid known regions of contamination for 1D power spectra.}
	\label{fig:2d_limit}
\end{figure}

Figure~\ref{fig:2d_limit} shows the residual 2D power spectra for the remaining 678 observations. We analyze the frequency band range 168.555--187.275\,MHz (approximately redshift 7) to avoid known instrumental effects (\S\ref{sec:AB2016}). The general features of the 2D power spectrum from Figure~\ref{fig:2d_AB_compare} are still present: foregrounds are most prevalent in the lowest $k_{||}$-mode and couple into the foreground wedge, channelizer flagging harmonics cause horizontal contamination lines, and there is lower power in the EoR window.

However, there are some key differences when comparing to the full 1029-observation integration in Figure~\ref{fig:2d_AB_compare}. The EoR window power is generally lower between the channelizer flagging harmonics, especially in the N--S polarization. A notable exception is the power at the lower left-hand corner of the EoR window; it remains fairly contaminated in the N--S polarization. The systematic that causes this contamination is unknown.

There is also noteworthy differences between the polarizations in Figure~\ref{fig:2d_limit}. The E--W polarization tends to have higher power than the N--S polarization in the EoR window, which will have significant consequences in 1D EoR upper limits. The cause for this is not yet known.

By understanding the regions affected by contamination, we can make logical cuts in \{$k_\perp,k_{||}$\}-space. This will allow us to calculate 1D EoR upper limits that are free of known contaminated $k$-modes.

\subsection{EoR upper limit}
\label{subsec:upper}

We can now perform informed \{$k_\perp,k_{||}$\}-cuts on the data to generate EoR upper limits. We avoid contaminated regions in Figure~\ref{fig:2d_limit} by making the following selections:
\begin{description}
    \item[$k_\perp$] Spatial modes outside of 18--80\,$\lambda$ are masked. This avoids low $k_\perp$-modes that are contaminated in the EoR window, and it removes any potential effects from poor $uv$-coverage at high $k_\perp$-modes.
    \item[$k_{||}$] We restrict line-of-sight modes to be greater than or equal to 0.15\,\textit{h}\,Mpc$^{-1}$. While the foreground wedge is mostly contained within the horizon line, there is some leakage at small $k_\perp$ into the EoR window, which is avoided with this mask.
    \item[$k_{||}$ vs. $k_\perp$] We add a small buffer to the horizon slope to avoid sub-horizon leakage, which is especially prevalent in the E--W polarization. Our slope is 15\% larger than the horizon, similar to the slopes used in \citet{beardsley_first_2016} and \citet{dillon_empirical_2015}.
\end{description}
Contours in Figure~\ref{fig:2d_limit} highlight the 2D region we use to generate 1D power spectra with these conditions.

\begin{figure*}[h]
\centering
	\includegraphics[width = .9\textwidth]{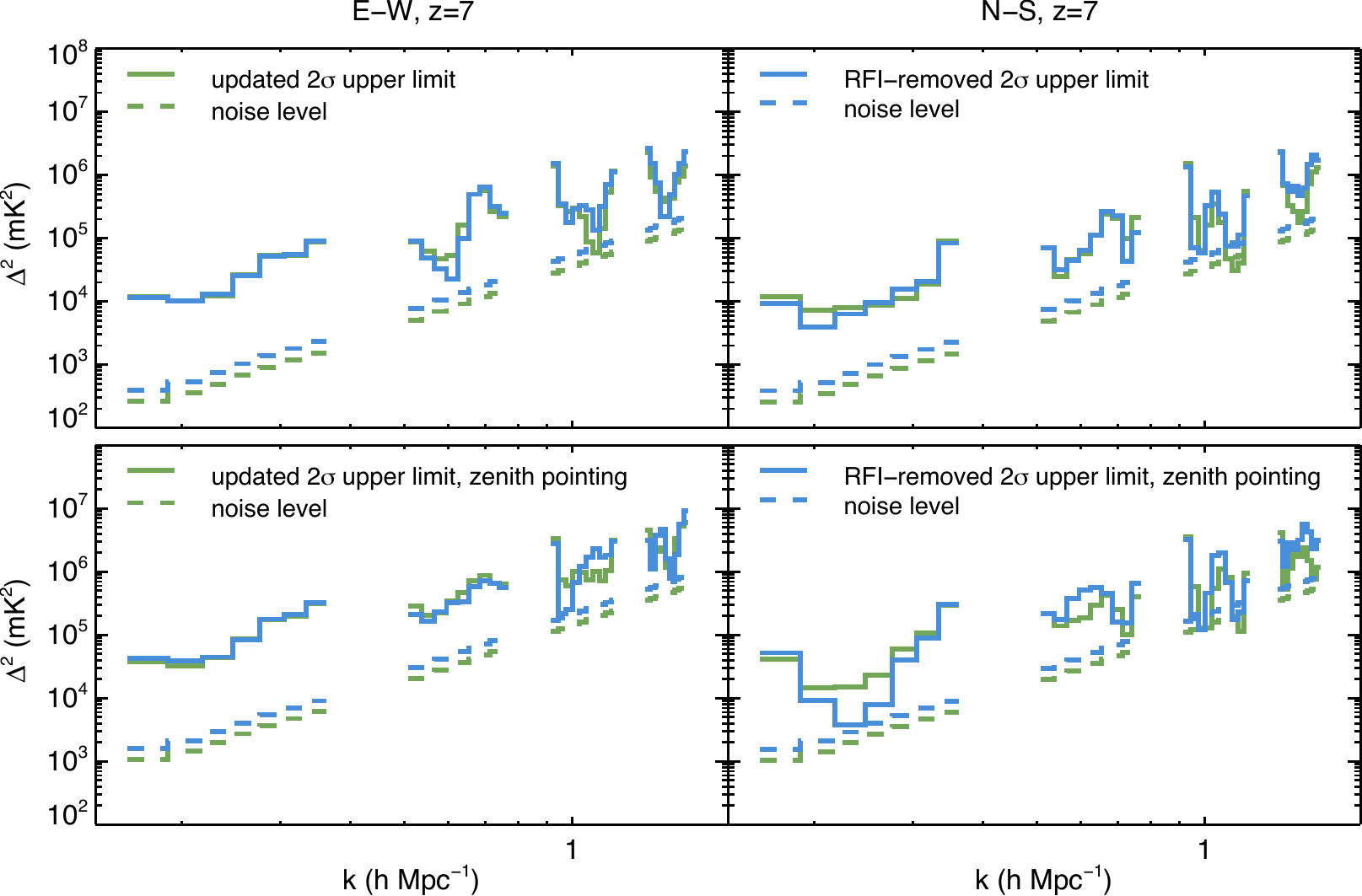}
	\caption{The 1D EoR upper limit comparison between our updated analysis for the 1029-observation data set (green) and our updated analysis for the RFI-removed 678-observation data set (blue) for the E--W and N--S polarizations at a band centered on redshift 7. There is marginal improvement over the full integration (top panel), however, there is significant improvement over the zenith-pointing subset (bottom panel). The dashed lines are the thermal noise levels of each analysis.}
	\label{fig:ssins_limit}
\end{figure*}

By choosing regions to integrate, we are potentially introducing a selection bias. This is unavoidable in a foreground-avoidance analysis; we must excise regions that we know are dominated by foregrounds in order to produce meaningful limits. To reduce the potential for selection bias, we only produce masks and cuts via foreground information from the 2D power spectrum. This lowers the degrees of freedom in our choice, and thus the potential bias.

Using these binning selections, we compare EoR upper limits from the full 1029-observation data set and the 678-observation subset selected with \textsc{ssins} in the top panel of Figure~\ref{fig:ssins_limit}. There is little to no change in the E--W polarization and a small improvement in the N--S polarization. However, the improvement via data selection from \textsc{ssins} is highly dependent on pointing. 

The bottom panel of Figure~\ref{fig:ssins_limit} shows the \textit{zenith pointing} of both the 1029-observation data set and the 678-observation subset. In this case, removing additional RFI using \textsc{ssins} improves the upper limit by a factor of 3.8 in the N--S polarization. Other pointings do not benefit from further RFI flagging despite similar RFI-detection levels, which indicates dominating systematics due to beam errors.

\begin{figure*}[h]
\centering
	\includegraphics[width = \textwidth]{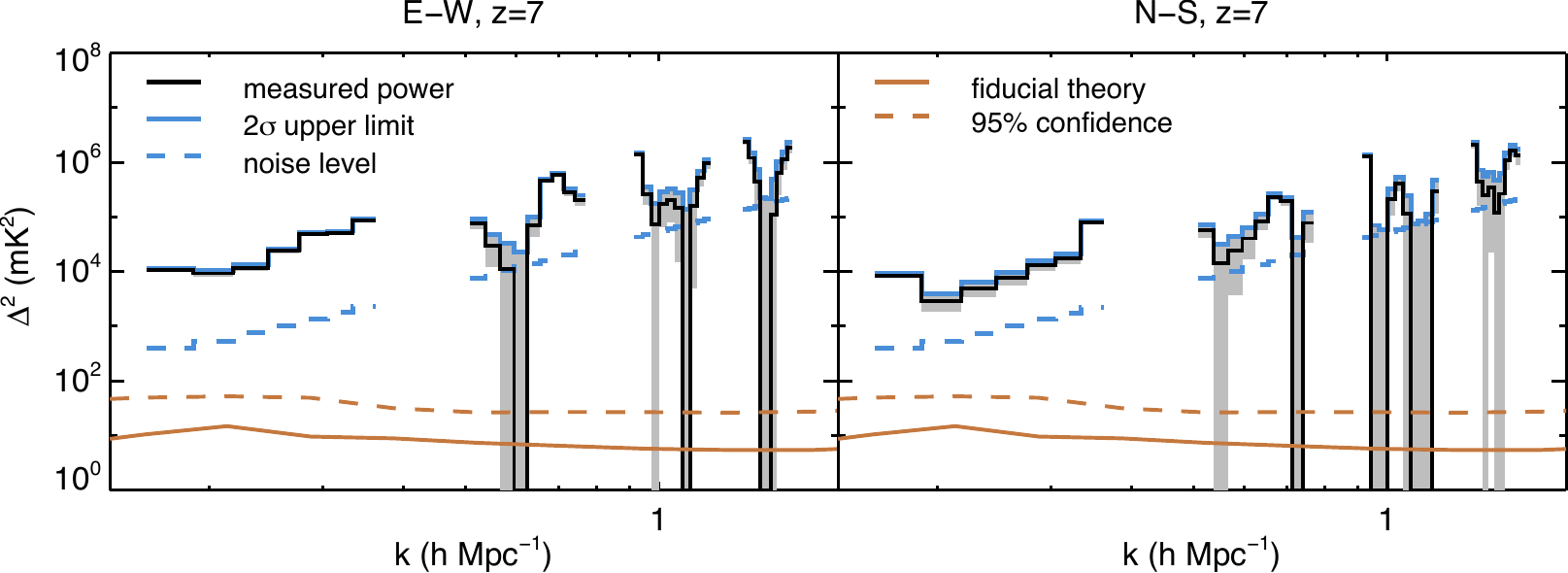}
	\caption{The 1D measured power spectra (black), the 2$\sigma$ error bars (gray), the 2$\sigma$ EoR upper limits (solid blue), and the 1$\sigma$ thermal noise levels (dashed blue) for the E--W and N--S polarizations using 678 observations selected with \textsc{ssins}. 
	 We also present an example fiducial EoR theory power spectrum (solid brown) along with the theoretical 2$\sigma$ upper limits on the 21\,cm power spectrum amplitude (brown dashed) obtained using existing observational constraints (see Appendix~\ref{simlimits} for further details).}
	These constitute our best EoR upper limits in this work.
	\label{fig:limit}
\end{figure*}

We present the 1D measured power spectra, the 2$\sigma$ error bars, the 2$\sigma$ EoR upper limits, and the 1$\sigma$ thermal noise levels in Figure~\ref{fig:limit}, along with an example EoR fiducial theory and its associated 2$\sigma$ upper limits (see Appendix~\ref{simlimits} for further details).

The measured power is the cross-power spectrum between the interleaved time samples (\S\ref{subsec:eppsilon}), and can fluctuate to be negative if below the thermal noise. The 1$\sigma$ noise level is the estimated noise from the integration using the fully propagated uncertainty estimates. The corresponding 2$\sigma$ error bars are shown for each power estimate, and can reach zero if the signal is consistent with a non-detection. The 2$\sigma$ EoR upper limit is calculated from the measured power and the variances, where a prior of being greater than the thermal noise is enforced. We omit bins that are severely affected by the channelizer-aliasing flagging. 

Our aim with this new integration is to be dominated by noise, not systematics. Since this is not enough data to theoretically detect the EoR, any detection is most likely that of a systematic in our data or analysis. Therefore, having a signal consistent with zero is ideal. We have many power estimates consistent with a non-detection, the majority of which are at higher $k$-modes. However, our lowest limits are systematic dominated. We can produce a similar, \textit{noise-dominated} limit ($\Delta^2 \leq 3.8 \times 10^3$\,mK$^2$ at $k=0.23$\,\textit{h}\,Mpc$^{-1}$) using just the zenith pointing. Given the observing approach of the MWA, it is not practical to use just zenith-pointing measurements of a designated field to detect the EoR in a timely fashion, therefore future efforts will need to reduce this systematic.

There are features in Figure~\ref{fig:limit} that indicate other sources of contamination via systematics. Firstly, the flagging from the channelizer aliasing causes foreground coupling on nearby $k$-modes. This indicates that a systematic floor might be present on \textit{all} $k$-modes at a lower level than this analysis can achieve, and thus will need to be addressed in the future. We know that only five modes in Figure~\ref{fig:limit} \textit{could} theoretically be below the 95\% confidence limit of the fiducial EoR by analyzing the power in the model, most likely due to this systematic. In addition, known cable reflections contaminate certain modes: 0.4\,\textit{h}\,Mpc$^{-1}$ (90\,m), 0.7\,\textit{h}\,Mpc$^{-1}$ (150\,m), 1.0\,\textit{h}\,Mpc$^{-1}$ (230\,m), 1.4\,\textit{h}\,Mpc$^{-1}$ (320\,m), 1.7\,\textit{h}\,Mpc$^{-1}$ (400\,m), and 2.3\,\textit{h}\,Mpc$^{-1}$ (524\,m). These modes cannot be used for EoR detection \citep{barry_calibration_2016}, and may contaminate surrounding modes at a low level \citep{ewall-wice_first_2016}.

The features present in the EoR upper limit, along with the features we cut via our mask and binning scheme, can help us determine future areas of improvement in our analysis and RFI-mitigation techniques (\S\ref{sec:future}).

\section{Data analysis validation}
\label{sec:checks}

Now that we have a new EoR power spectrum upper limit, we present various procedures to prove its validity. Specifically, we compare results with another pipeline, perform a signal loss simulation, and present error propagation results from {\eppsilon}.

\subsection{Pipeline comparison}
\label{subsec:pipe_compare}

We cross-validate our FHD/{\eppsilon} pipeline results with a different set of packages: the RTS (Real Time System, \citealt{mitchell_real-time_2008,ord_interferometric_2010}) and CHIPS (Cosmological H\,I Power Spectrum, \citealt{trott_chips:_2016}) pipeline. We have extensively used this verification approach in both \citet{jacobs_murchison_2016} and \citet{beardsley_first_2016} to indicate proper treatment of normalization, polarization, and signal preservation. By comparing power spectra from the same data set, we can demonstrate the continued robustness of our analysis pipelines.

The RTS calibrates raw data and produces visibilities, therefore it plays a similar role to FHD. However, the methodology is quite different. Specifically, the RTS can perform direction-dependent calibration via a peeling method. Approximately 1000 sources are used for a preliminary direction-independent calibration, and then five bright sources are used to estimate the specific calibration in the local region. Separate calibration values are calculated per 1.28\,MHz (per coarse band). This constitutes the largest philosophical difference, but a variety of processes are distinct, including other calibration parameters, beam calculations, and techniques to enforce spectral smoothness. Given the importance of these features in the power spectrum space (\S\ref{subsec:FHD}), comparisons with the RTS are relevant for validation.

The RTS-calibrated visibilities are then processed by CHIPS, which starts by gridding the visibilities onto the $\{u,v,w\}$-plane. By choosing discrete $w$-projection planes, CHIPS avoids the need to use image space for integrating observations. Therefore, many of the required aliasing mitigation techniques discussed in \S\ref{subsec:FHD} are not necessary with this package. CHIPS also uses advanced systematic mitigation techniques, including an inverse-covariance weighting scheme during power estimation \citep{kay_fundamentals}. The noise calculation used for this comparison is a propagated noise from the even--odd difference, similar to the approach in {\eppsilon}.

\begin{figure*}
\centering
	\includegraphics[width = \textwidth]{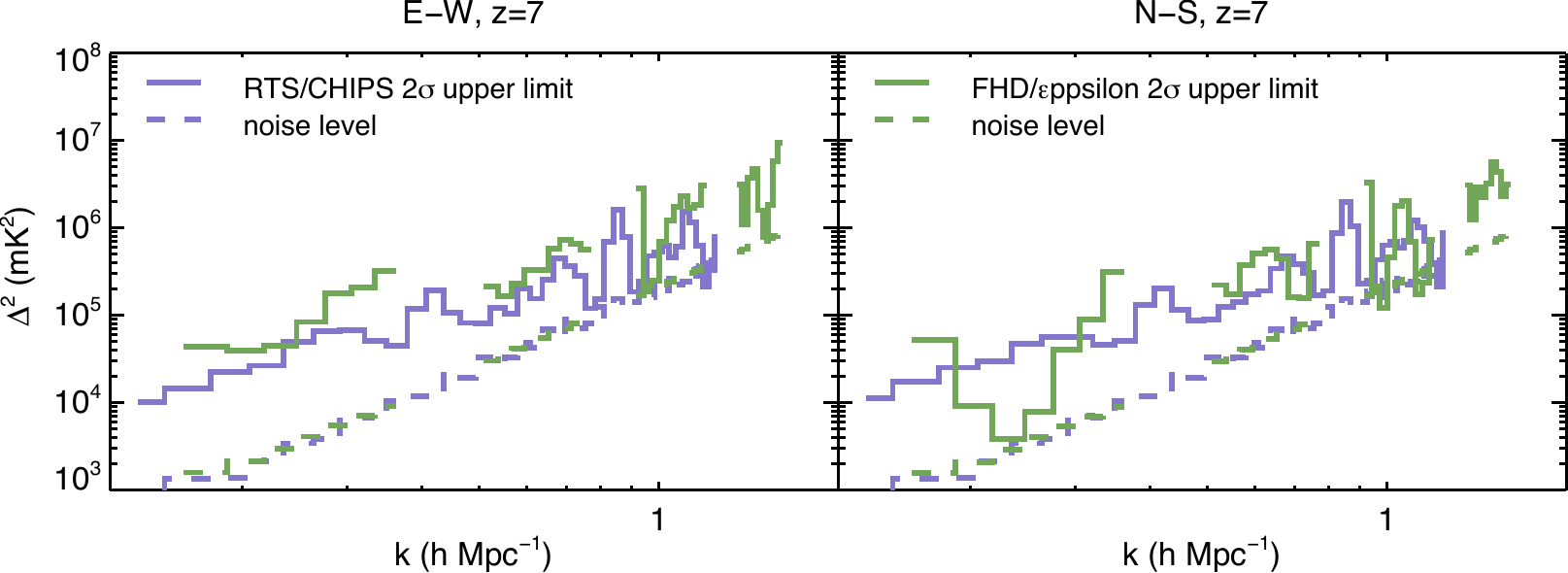}
	\caption{A cross-validation analysis of the EoR upper limits and associated noise levels on the zenith-pointing subset from the FHD/{\eppsilon} pipeline (green) and the RTS/CHIPS pipeline (purple). RTS/CHIPS recovers more $k$-modes in known systematic-dominated regions via advanced techniques, while FHD/{\eppsilon} produces lower systematics at some low $k$-modes. In general, their consistency with each other demonstrates robustness in our analysis techniques.}
	\label{fig:Aug23_compare}
\end{figure*}

We compare the zenith-pointing subset for our cross-validation in Figure~\ref{fig:Aug23_compare}. To remain consistent, we apply the same binning scheme in \S\ref{subsec:upper} to each data reduction. This foreground/systematic avoidance scheme was chosen with the FHD/{\eppsilon} analysis in mind; they may not be the best cuts for RTS/CHIPS. Nevertheless, the EoR upper limits are roughly consistent with each analysis, and follow a general trend within the same order of magnitude for all $k$. 

Since CHIPS uses an inverse-covariance weighting, the RTS/CHIPS analysis recovers more $k$-modes in the channelizer-aliasing harmonics. This technique also allows the RTS/CHIPS analysis to probe further within the foreground wedge, and thus their lowest limit is at their lowest $k$-mode. In contrast, the FHD/{\eppsilon} analysis has lower systematics between the flagging harmonics. 

In general, the two analyses produce similar EoR upper limits from the same data set. While 5\,hr is not enough to create a competitive limit, we show continued consistency between two unique pipelines.

\subsection{Pipeline simulation}
\label{subsec:pipe_sim}

In addition to our analysis cross-validation with RTS/CHIPS, we simulate our pipeline end-to-end to test for signal loss. This demonstrates self-consistency and signal preservation throughout the analysis.

In order to validate FHD specifically, we run an in-situ simulation. FHD is naturally an instrument simulator; we produce model visibilities that represent the response of the instrument to tens of thousands of point sources on the sky. These model visibilities, plus a theoretical EoR response, can be input into FHD as simulation data. 

We run two pipeline-verification tests: (1) theoretical EoR signal through the entire pipeline with no calibration or subtraction and (2) theoretical EoR in a point-source sky where only a subset of the brightest point sources are subtracted with no calibration effects. These two tests investigate whether we recover the input EoR signal and if we can theoretically detect the EoR with realistic foregrounds disregarding calibration effects, respectively.

\begin{figure}
\centering
	\includegraphics[width=\columnwidth]{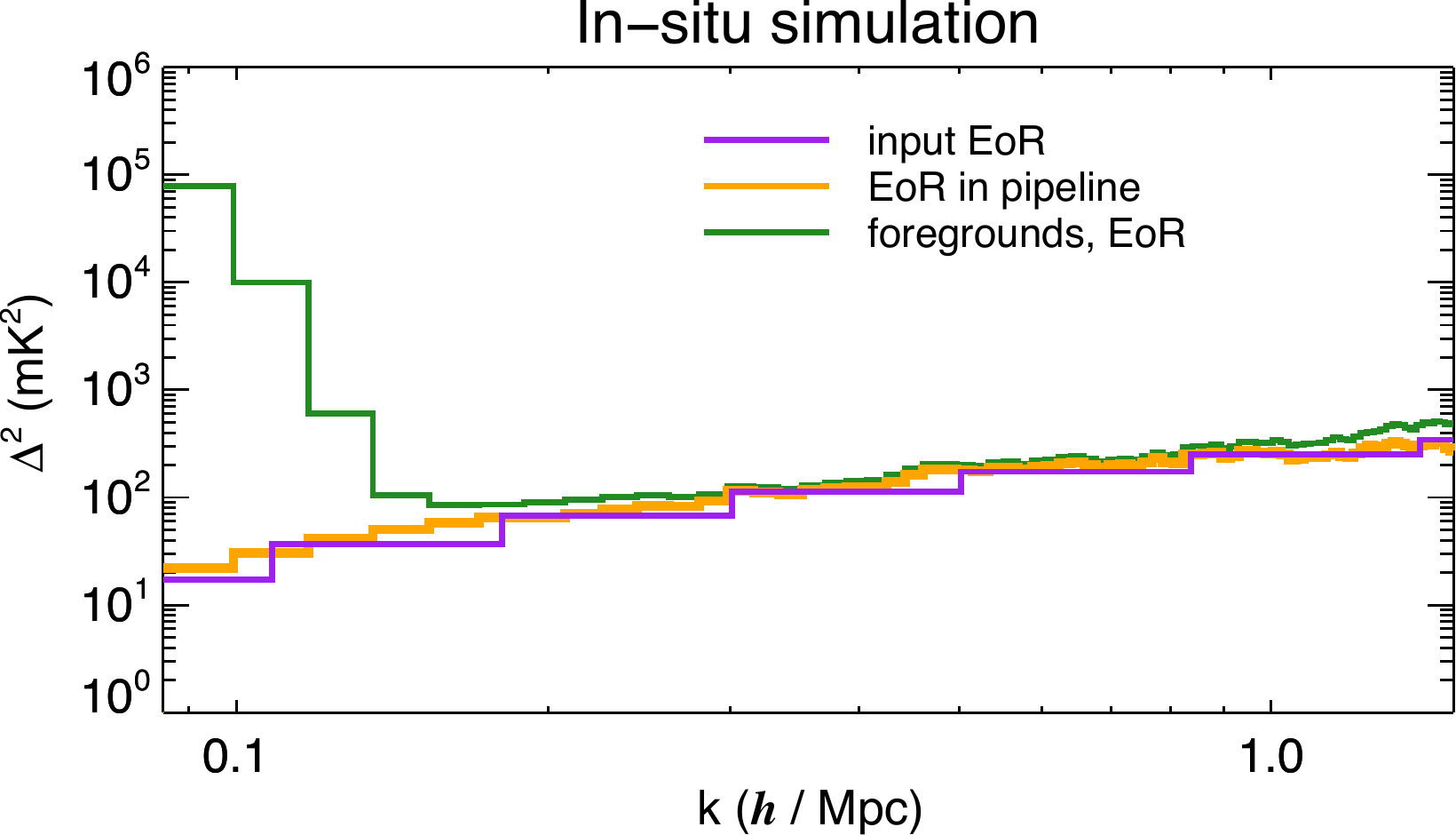}
	\caption{Measured 1D power for two in-situ simulations on the FHD/{\eppsilon} pipeline. We input simulated EoR visibilities (purple) into the pipeline and recover the expected power (orange). If we add foregrounds and only subtract a subset (green), we still recover the underlying EoR signal for most all $k$-modes.}
	\label{fig:signal_loss}
\end{figure}

Figure~\ref{fig:signal_loss} shows the measured power from these two simulations in 1D power spectrum space. We include the foreground wedge and the full bandwidth in this binning scheme. Our calculated power from the input EoR visibilities indicates that we do not suffer from signal loss or normalization errors; we recover the input power across all $k$-modes. When there are residual foregrounds in the simulation, we still recover the input EoR simulation in the EoR window for a large range of $k$-modes. Large $k$-modes (above $\sim1.0$\,\textit{h}\,Mpc$^{-1}$) are coupled to foreground contamination via our gridding kernel resolution \citep{beardsley_first_2016}.

Therefore, we have verified the ability of FHD to detect the EoR in ideal conditions without calibration effects for a wide range of $k$-modes. We report the results of these types of simulations whenever we reanalyze data with updated software, therefore similar simulations were performed in \citet{barry_calibration_2016} and \citet{barry_fhd/eppsilon_2019} on the FHD/{\eppsilon} pipeline.

\subsection{Uncertainty estimation verification}
\label{subsec:uncertainty}

We also conduct verification experiments within {\eppsilon} to test the validity of our noise assumptions. Given that our best EoR upper limit is noise dominated, it is crucial to investigate our uncertainty measurements. 

First, we can determine the level of contamination in the noise due to choosing image space as our integration basis. While noiseless pipeline simulations have demonstrated that our measured power estimation recovers the EoR in \S\ref{subsec:pipe_sim}, the noise can still be adversely affected by image-space aliasing. 

We can calculate the noise contamination level by comparing the analysis of one observation using various integration schemes. By only analyzing one observation, it is not necessary to go to image space because integration is not required. Therefore, we have the capability to compare to noise that is not affected by the transform to image space. 

\begin{figure}
\centering
	\includegraphics[width=\columnwidth]{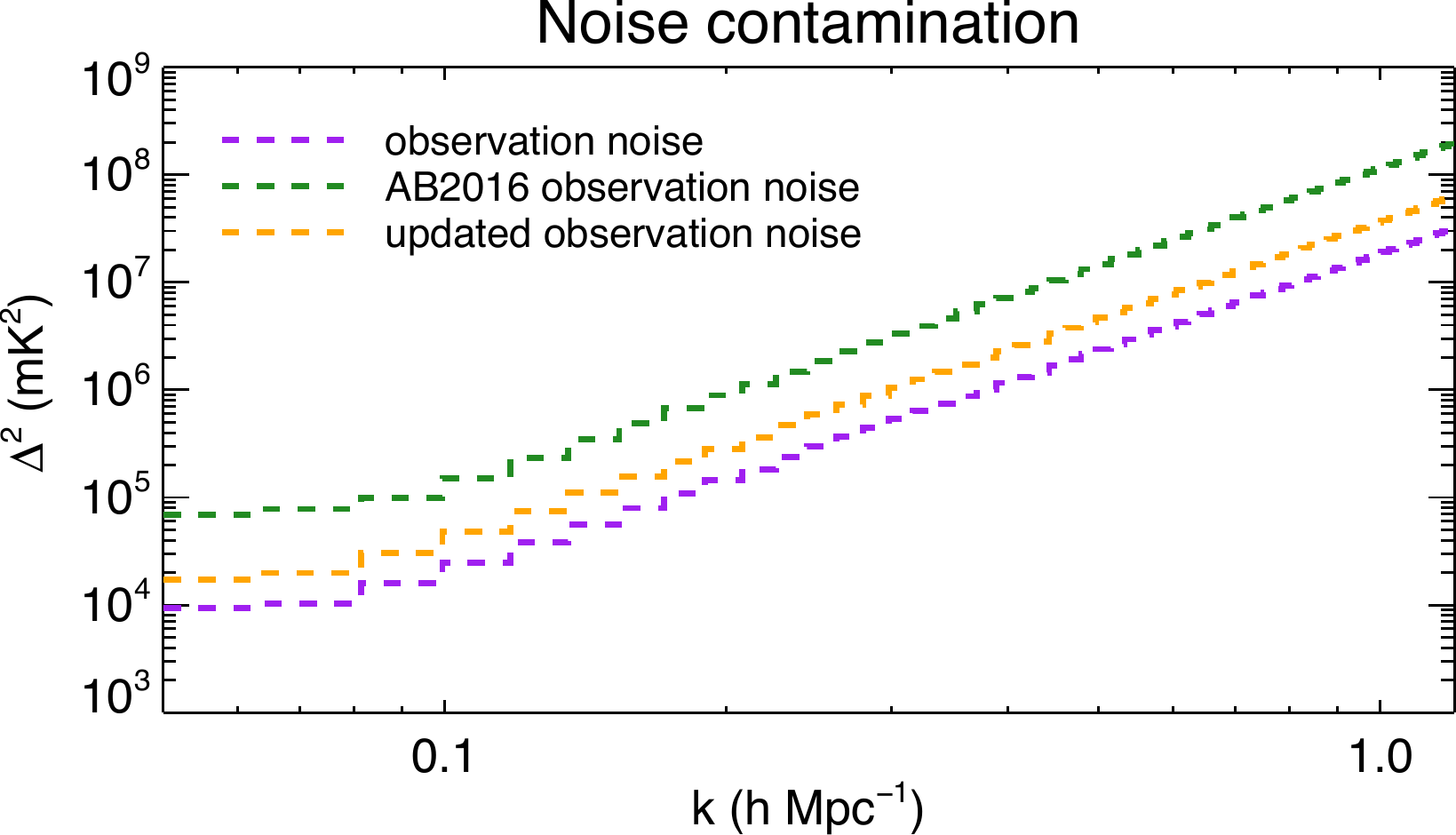}
	\caption{Calculated thermal noise levels for one observation using various integration schemes. Avoiding image space results in the lowest noise level (purple). Using a small image to match \citet{beardsley_first_2016} results in a factor of 6 contamination (green), and using our updated method results in a factor of 2 contamination (orange). Mitigating this effect is left for future work.}
	\label{fig:noise_obs}
\end{figure}

In this work, we have two varieties of HEALPix integration schemes. \citet{beardsley_first_2016} used a square $\sim$20$\degree$ region, while our updated analysis uses a region that includes more than 10$\times$ the area and includes a modified gridding kernel. These two tests, along with the noise level calculated without an image-space transform, are shown in Figure~\ref{fig:noise_obs}.

The excess contamination from the image-space transform is flat across all $k$-modes. The integration scheme from \citet{beardsley_first_2016} is high by about a factor of 6, whereas our updated integration scheme is high by about a factor of 2. While this is an improvement, we need to mitigate this contamination in the future. We \textit{do not} apply a correction factor in this work, therefore our EoR upper limit is higher than it could theoretically be for our analysis. 

We also analytically calculate the error propagation throughout {\eppsilon}, thereby avoiding the ambiguity of bootstrapped errors. However, in order to make this manageable, we assume there are no cross-correlations between pixels in $\{u,v,f\}$-space even though there are various stages in the analysis that could cause correlations. Increasing the integrated image size creates smaller $uv$-pixels, which are more likely to be correlated. Applying the modified gridding kernel and the frequency window will also correlate pixels. 

To investigate the assumption of independent pixels in our analytic calculation, we also measure the standard deviation of the noise power created from the even--odd difference. These uncertainties are noisy; they will be subject to random noise variations per pixel. This gives us two different noise estimates: (1) an observed noise calculated from the power, and (2) an expected noise calculated from error propagation of the input cubes. The observed noise and expected noise should be the same magnitude if our error propagation and correlation assumptions are well-founded.

\begin{figure*}[h!]
\centering
	\includegraphics[width = .75 \textwidth]{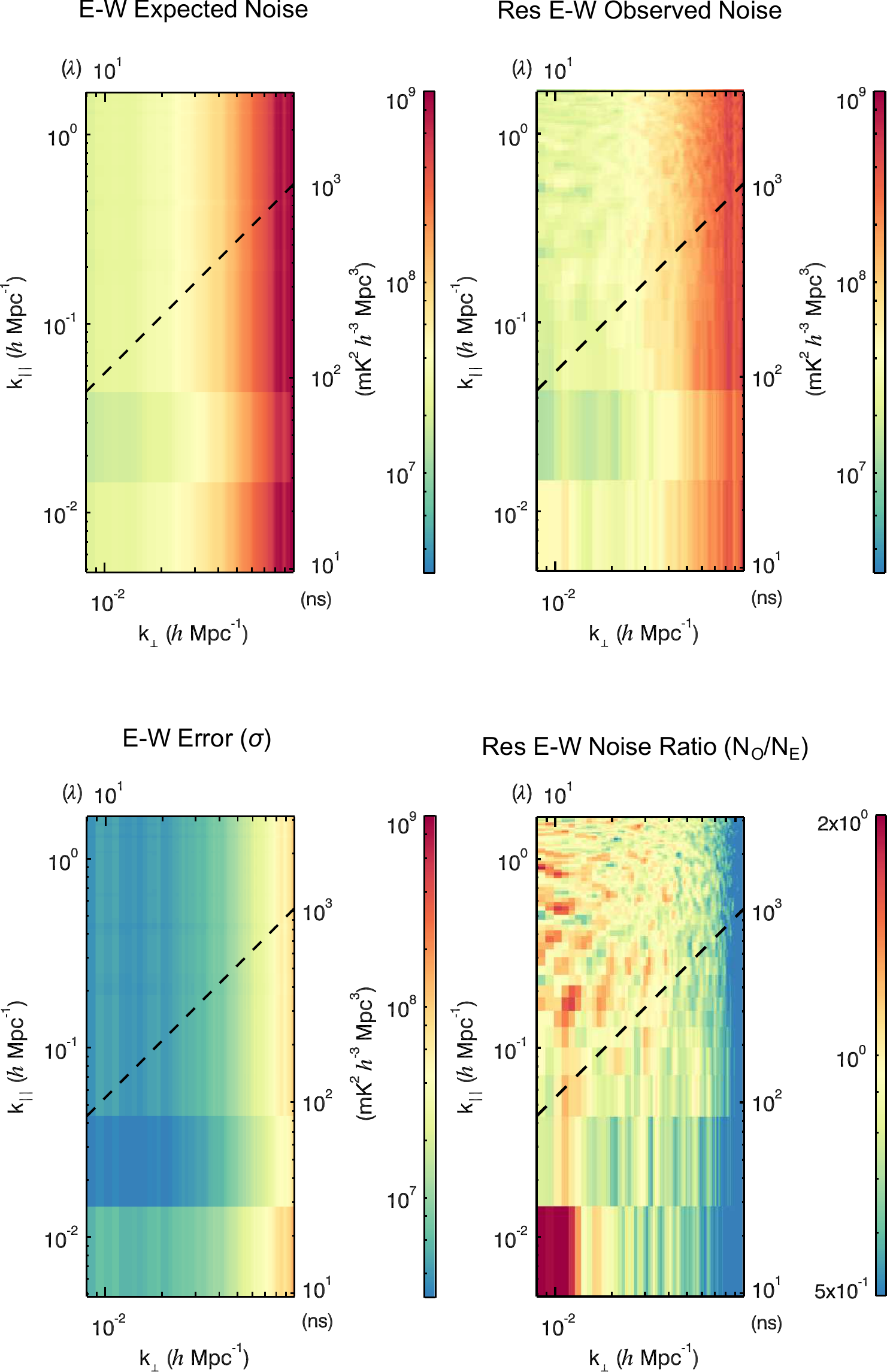}
	\caption{The observed noise from the resulting power (top right), the expected noise from the analytic uncertainty estimate (top left), the resulting analytic error bars (bottom left), and the ratio between the expected and observed noise (bottom right). We validate our analytic error propagation and assumptions in {\eppsilon} via the noise ratio, which is very close to 1.}
	\label{fig:error_panel}
\end{figure*}

Figure~\ref{fig:error_panel} shows the noise and error 2D power spectra calculated from the 678-observation data set in \S\ref{subsec:upper}. The observed noise (top right) and the expected noise (top left) are very similar, and their ratio (bottom right) is very close to 1. There is some deviation from 1 in poor $uv$-coverage regions, however these will not contribute to the 1D power spectrum given our binning schemes. The error bars (bottom left) are related to the expected noise. 

By comparing the observed noise to the expected noise, we have investigated whether cross-correlations caused by our analysis techniques have a significant effect. A ratio close to unity indicates no excessive cross-correlations. 

\section{Discussion}
\label{sec:future}

Our new EoR upper limit with the open-source FHD/{\eppsilon} pipeline is almost an order of magnitude better than previous data reductions, and there have been many contributing factors to this improvement. These developments can be classified under four main modifications: (1) change in flux density scale, (2) change in analysis that reduced the measured power spectrum value, (3) change in analysis that reduced the contamination in the noise, and (4) change in RFI mitigation. 

Adopting a new catalog for better calibration and subtraction accuracy lowered our power spectrum normalization by approximately 1.3. Improving our analysis through the modified gridding kernel and other various techniques reduced contamination in both the measured power and the noise for a combined reduction of 2.8 in power. Finally, excising observations contaminated with faint RFI lowered our limit by a factor of 3.8, but only for the zenith pointing. This has highlighted the critical aspects of precision data analysis on the EoR, and indicates areas that we can continue to develop.

Future data reductions with FHD/{\eppsilon} can further improve the EoR upper limit by enforcing spectral smoothness in systematic errors. This encompasses a large breadth of error types, including those from the theoretical beam kernel, sky calibration, and HEALPix interpolation. However, some frequency-dependent errors are inherent to the analysis, like the discreteness of the $uv$-plane during the estimation of the model visibilities and the gridding process \citep{kerrigan_improved_2018}.

This has culminated in a new general approach in our analysis. We sacrifice modes \textit{not} within the EoR window in order to keep modes in sensitive measurement regions as spectrally precise as possible. For example, the tapered gridded estimator that we apply to the beam kernel when we calculate power spectra is not an accurate representation of the sky. Our subtraction model visibilities will be dominated from sources in the center of the beam, and sources near the horizon will be significantly down-weighted. In order to recover modes in the foreground wedge, sidelobe sources must be included and be precise \citep{pober_importance_2016}. However, the modified gridding kernel reduces the spectral dependence of the discrete-based errors in the EoR window. Thus, we knowingly do not recover modes within the foreground wedge in order to reduce errors in the EoR window.

In the future, we plan to investigate other methods to reduce spectral errors which affect the sensitive measurement modes, specifically those based on the instrument (e.g. the beam and calibration). However, this is difficult if the actual response is spectrally complicated. It is easier for the errors we make to be spectrally smooth if the response is \textit{also} spectrally smooth. This will advise upgrades to the MWA and the design of the Square Kilometre Array (SKA). 

Comparisons with other data pipelines have also demonstrated potential ways to remove systematics from more $k$-modes. The RTS/CHIPS pipeline uses inverse-covariance weighting and other methods to help remove systematics caused from flagging channelizer aliasing. We are dominated on some $k$-modes from this effect in the FHD/{\eppsilon} pipeline, with potential for it to affect all $k$-modes at a low power level. Incorporating these advanced techniques into power spectrum estimation will be an important aspect of lowering the EoR upper limit in our data reductions.

One of the major improvements to our EoR upper limit has come from using the package \textsc{ssins} to remove RFI-contaminated observations from our integrations. We are beginning to enter an analysis regime where low-level contamination can dominate the power in the EoR window. In the future, we plan to use the \textsc{ssins} package to its fullest potential in order to recover observations with faint DTV while still maintaining a high degree of RFI-clean observations. Nevertheless, we have only gained significant improvements from further RFI flagging in the zenith pointing of the N--S polarization. This indicates that there is significant beam errors in non-zenith pointings and some residual systematic in the E--W polarization. Mitigating these contaminations will be investigated in the future. 

To conclude our work, we put our results in the context of the wider EoR community. Previous publications from the MWA report $\Delta^2 \leq 2.7 \times 10^4$\,mK$^2$ at $k = 0.27$\,\textit{h}\,Mpc$^{-1}$ and $z = 7.1$ \citep{beardsley_first_2016}. The PAPER Collaboration, who previously had the lowest EoR upper limits \citep{ali_paper-64_2015}, has recently reanalyzed these calculations due to signal loss \citep{cheng_characterizing_2018, kolopanis_simplified_2019}. The next lowest current limits come from the LOFAR Collaboration, with $\Delta^2 \leq 6.3 \times 10^3$\,mK$^2$ at $k = 0.053$\,\textit{h}\,Mpc$^{-1}$ and $z = 10.1$ \citep{patil_upper_2017}. 

Our best limit from this work is $\Delta^2 \leq 3.9 \times 10^3$\,mK$^2$ at $k=0.20$\,\textit{h}\,Mpc$^{-1}$ and $z=7$ in the N--S polarization on 21\,hr. This is currently the lowest upper limit on EoR structure in the literature. We note that all detections in this work are those of systematics, and longer integrations are needed to reduce thermal noise to the level of the EoR.

We report all $k$-modes and associated limit calculations from this work in Appendix~\ref{appendix} for ease of comparison. In addition, both the data\footnote{Data available via MWA All-Sky Virtual Observatory portal (\url{http://www.mwatelescope.org/data}) and a public Amazon Web Services mirror.} and the software\footnote{Software is open source and freely available as part of the EoRImaging GitHub repository (\url{https://github.com/EoRImaging}).} is publicly available. The full list of observations, software versions, and settings files necessary to recreate this analysis are available upon request.

By incorporating pipeline improvements to reduce analysis systematics and by removing RFI contamination to reduce observational systematics, we are now in the regime where we are noise dominated in our lowest limit. This puts emphasis on both fronts: lowering the EoR upper limit will require development in analysis precision and observational contamination mitigation. We have proven that both are crucial for detecting the structure of the EoR in the power spectrum.


\acknowledgments

This research was supported by the Australian Research Council Centre of Excellence for All Sky Astrophysics in 3 Dimensions (ASTRO 3D), through project number CE170100013. This work was supported by resources awarded under Astronomy Australia Ltd’s merit allocation scheme on the OzSTAR national facility at Swinburne University of Technology. OzSTAR is funded by Swinburne University of Technology and the National Collaborative Research Infrastructure Strategy (NCRIS). APB is supported by an NSF Astronomy and Astrophysics Postdoctoral Fellowship under \#1701440. CMT is supported by an ARC Future Fellowship under grant FT180100196. This work has been funded by National Science Foundation grants AST-1410484, AST-1506024, AST-1613040, AST-1613855, AST-1643011, and AST-1835421, and Amazon/SKA Astrocompute led by D. Jacobs at Arizona State. The International Centre for Radio Astronomy Research (ICRAR) is a Joint Venture of Curtin University and The University of Western Australia, funded by the Western Australian State government. This scientific work makes use of the Murchison Radio-astronomy Observatory, operated by CSIRO. We acknowledge the Wajarri Yamatji people as the traditional owners of the Observatory site. Support for the operation of the MWA is provided by the Australian Government (NCRIS), under a contract to Curtin University administered by Astronomy Australia Limited. We acknowledge the Pawsey Supercomputing Centre which is supported by the Western Australian and Australian Governments.

%

\vspace{5mm}
\facilities{MWA}





\appendix

\section{Theoretical upper limits on the 21cm power spectrum} \label{simlimits}

To construct the fiducial theory 21cm power spectrum used in Figure~\ref{fig:limit} (brown solid curve) we use \textsc{\small 21cmFAST} \citep{Mesinger_21cmFAST_2011}, adopting the latest astrophysical model parameterization from \citet{park_2019} and assume that we are in the saturated limit (the 21cm spin temperature is considerably larger than the CMB temperature). 

Rather than using just a single fiducial model, we can also explore the allowed variation in the amplitude of the theoretical 21cm power spectrum owing to the uncertainties in the underlying reionization astrophysics. We achieve this by calculating theoretical 2$\sigma$ upper limits on the 21cm power spectrum model (brown dashed curve in Figure~\ref{fig:limit}) following a similar approach to that of \citet{pober_limits_2016}. That is, we use \textsc{\small 21CMMC} \citep{Greig_2015}, a Monte-Carlo Markov-Chain sampler of 3D reionization simulations, and use existing observational constraints on the reionization epoch to constrain the theoretical astrophysical models. Specifically, we only consider the six astrophysical parameters governing the ionizing sources from \citet{park_2019} and assume the spin temperature is saturated. These six astrophysical parameters include both a mass dependent escape fraction and fraction of gas in stars, a minimum turn-over mass for haloes hosting star-forming galaxies and a star-formation time-scale. Additionally, we ignore recombinations, and thus include a mean photon horizon, $R_{\rm mfp}$ parameter.

For our observational constraints, we use the limits on the intergalactic medium neutral fraction at the tail-end of reionization (at $z=5.9$; $x_{\rm H{\scriptstyle I}} < 0.06 + 0.05~(1\sigma)$) from the dark pixel statistics from quasar spectroscopy \citep{McGreer_2015}, the latest estimate for the electron scattering optical depth from Planck ($\tau=0.054\pm0.007$, \citealt{Planck_2018}) and the ultra-violet galaxy luminosity functions at $z=6, 7, 8$~and~10 \citep{Bouwens_2015,Bouwens_2017,Oesch_2018}. Post-processing the output 21cm power spectrum data from each sampled astrophysical model from the MCMC, we can construct marginalized probability distribution functions (PDFs) for the power spectrum amplitude as a function of Fourier mode, $k$. Our theoretical limit on the 21cm power spectrum is then obtained from sampling these PDFs.

\section{All calculated EoR upper limits}
\label{appendix}

We report all EoR upper limits from this work to aid the community in creating comparisons. This data set consists of 678 observations, whose selection from the full 1029 observation data set is discussed in \S\ref{subsec:data_select}. The various masks in our binning scheme are presented in \S\ref{subsec:upper}, and we choose to analyze the frequency range ${\small \sim}$168.5\,MHz -- 187.3\,MHz to avoid known instrumental effects. 

The E--W values (left) and N--S values (right) are reported in Table~\ref{table:all_eor}. We include the $k$ (\textit{h}\,Mpc$^{-1}$), the upper limit $\Delta_U^2$ (mK$^2$), the lower uncertainty bound $\Delta_L^2$ (mK$^2$), the measured power $\Delta^2$ (mK$^2$), and the 1$\sigma$ thermal noise (mK$^2$). If the $k$-mode is consistent with a non-detection (e.g. the measured power or the calculated lower uncertainty bound is negative) we report a lower uncertainty bound of zero.

\renewcommand{\arraystretch}{1.1}
\begin{table}{}
\centering

\begin{tabular}{||c c c c c||c c c c c||} 
\hline
\multicolumn{5}{||c||}{E--W} & \multicolumn{5}{c||}{N--S} \\
\hline
$k$ & $\Delta_U^2$ & $\Delta_L^2$ & $\Delta^2$ & Thermal & $k$ & $\Delta_U^2$ & $\Delta_L^2$ & $\Delta^2$ & Thermal \\
\hline\hline

0.174	&	1.13$\times 10^	4	$ &	9.75$\times 10^	3	$ &	1.05$\times 10^	4	$ &	3.98$\times 10^	2$	& 0.174	&	9.22$\times 10^	3	$ &	7.67$\times 10^	3	$ &	8.45$\times 10^	3	$ &	3.87$\times 10^	2	$	\\ \hline
0.203	&	1.03$\times 10^	4	$ &	8.14$\times 10^	3	$ &	9.21$\times 10^	3	$ &	5.35$\times 10^	2$	& 0.203	&	3.89$\times 10^	3	$ &	1.81$\times 10^	3	$ &	2.85$\times 10^	3	$ &	5.21$\times 10^	2	$   \\ \hline
0.232	&	1.31$\times 10^	4	$ &	1.02$\times 10^	4	$ &	1.16$\times 10^	4	$ &	7.46$\times 10^	2$	& 0.232	&	6.32$\times 10^	3	$ &	3.42$\times 10^	3	$ &	4.87$\times 10^	3	$ &	7.25$\times 10^	2	$	\\ \hline
0.261	&	2.58$\times 10^	4	$ &	2.17$\times 10^	4	$ &	2.38$\times 10^	4	$ &	1.03$\times 10^	3$	& 0.261	&	9.61$\times 10^	3	$ & 5.61$\times 10^	3	$ &	7.61$\times 10^	3	$ &	9.99$\times 10^	2	$	\\ \hline
0.290	&	5.19$\times 10^	4	$ &	4.64$\times 10^	4	$ &	4.92$\times 10^	4	$ &	1.37$\times 10^	3$	& 0.290	&	1.56$\times 10^	4	$ &	1.03$\times 10^	4	$ &	1.29$\times 10^	4	$ &	1.33$\times 10^	3	$	\\ \hline
0.319	&	5.47$\times 10^	4	$ &	4.75$\times 10^	4	$ &	5.11$\times 10^	4	$ &	1.79$\times 10^	3$	& 0.319	&	2.07$\times 10^	4	$ &	1.37$\times 10^	4	$ &	1.72$\times 10^	4	$ &	1.74$\times 10^	3	$	\\ \hline
0.349	&	8.99$\times 10^	4	$ &	8.07$\times 10^	4	$ &	8.53$\times 10^	4	$ &	2.30$\times 10^	3$	& 0.349	&	8.32$\times 10^	4	$ &	7.42$\times 10^	4	$ &	7.87$\times 10^	4	$ &	2.23$\times 10^	3	$	\\ \hline

0.523	&	9.11$\times 10^	4	$ &	6.06$\times 10^	4	$ &	7.58$\times 10^	4	$ &	7.63$\times 10^	3$	& 0.523	&	7.09$\times 10^	4	$ &	4.13$\times 10^	4	$ &	5.61$\times 10^	4	$ &	7.40$\times 10^	3	$	\\ \hline
0.552	&	4.81$\times 10^	4	$ &	1.22$\times 10^	4	$ &	3.02$\times 10^	4	$ &	8.97$\times 10^	3$	& 0.552	&	3.20$\times 10^	4	$ &	0                     &	1.44$\times 10^	4	$ &	8.71$\times 10^	3	$	\\ \hline
0.581	&	3.25$\times 10^	4	$ & 0  $                $ &	1.09$\times 10^4	$ &	1.05$\times 10^	4$	& 0.581	&	4.43$\times 10^	4	$ &	3.60$\times 10^	3	$ &	2.39$\times 10^	4	$ &	1.02$\times 10^	4	$	\\ \hline
0.610	&	2.26$\times 10^	4	$ & 0  $                $ &	-8.02$\times 10^3	$ &	1.21$\times 10^	4$	& 0.610	&	6.38$\times 10^	4	$ &	1.69$\times 10^	4	$ &	4.04$\times 10^	4	$ &	1.17$\times 10^	4	$	\\ \hline
0.639	&	9.88$\times 10^	4	$ &	4.35$\times 10^	4	$ &	7.12$\times 10^	4	$ &	1.38$\times 10^	4$	& 0.639	&	1.11$\times 10^	5	$ &	5.70$\times 10^	4	$ &	8.38$\times 10^	4	$ &	1.34$\times 10^	4	$	\\ \hline
0.668	&	4.94$\times 10^	5	$ &	4.31$\times 10^	5	$ &	4.63$\times 10^	5	$ &	1.59$\times 10^	4$	& 0.668	&	2.62$\times 10^	5	$ &	2.00$\times 10^	5	$ &	2.31$\times 10^	5	$ &	1.54$\times 10^	4	$	\\ \hline
0.697	&	6.36$\times 10^	5	$ &	5.64$\times 10^	5	$ &	6.00$\times 10^	5	$ &	1.81$\times 10^	4$	& 0.697	&	2.28$\times 10^	5	$ &	1.58$\times 10^	5	$ & 1.93$\times 10^	5	$ &	1.75$\times 10^	4	$	\\ \hline
0.726	&	3.20$\times 10^	5	$ &	2.38$\times 10^	5	$ &	2.79$\times 10^	5	$ &	2.04$\times 10^	4$	& 0.726	&	4.25$\times 10^	4	$ &	0               	  &	-3.93$\times 10^3	$ &	1.98$\times 10^	4	$	\\ \hline
0.755	&	2.49$\times 10^	5	$ &	1.57$\times 10^	5	$ &	2.03$\times 10^	5	$ &	2.30$\times 10^	4$	& 0.755	&	1.22$\times 10^	5	$ &	3.29$\times 10^	4	$ &	7.75$\times 10^	4	$ &	2.23$\times 10^	4	$	\\ \hline

0.929	&	1.50$\times 10^	6	$ &	1.33$\times 10^	6	$ &	1.41$\times 10^	6	$ &	4.29$\times 10^	4$	& 0.929	&	1.35$\times 10^	6	$ &	1.19$\times 10^	6	$ &	1.27$\times 10^	6	$ &	4.16$\times 10^	4	$   \\ \hline
0.958	&	3.52$\times 10^	5	$ &	1.64$\times 10^	5	$ &	2.58$\times 10^5	$ &	4.70$\times 10^	4$	& 0.958	&	7.01$\times 10^	4	$ &	0                     &	-6.04$\times 10^4	$ &	4.56$\times 10^	4	$   \\ \hline
0.987	&	1.77$\times 10^	5	$ &	0                     &	7.25$\times 10^4	$ &	5.14$\times 10^	4$	& 0.987	&	5.91$\times 10^	4	$ &	0                     &	-1.15$\times 10^5	$ &	4.99$\times 10^	4	$	\\ \hline
1.017	&	2.88$\times 10^	5	$ &	6.33$\times 10^	4	$ &	1.75$\times 10^5	$ &	5.61$\times 10^	4$	& 1.017	&	3.24$\times 10^	5	$ &	1.06$\times 10^	5	$ &	2.15$\times 10^5	$ &	5.44$\times 10^	4	$	\\ \hline
1.046	&	3.24$\times 10^	5	$ &	8.01$\times 10^	4   $ &	2.02$\times 10^5	$ &	6.09$\times 10^	4$	& 1.046	&	5.32$\times 10^	5	$ &	2.96$\times 10^	5	$ &	4.14$\times 10^	5	$ &	5.91$\times 10^	4	$   \\ \hline
1.075	&	2.78$\times 10^	5	$ &	1.48$\times 10^	4   $ &	1.46$\times 10^	5	$ &	6.57$\times 10^	4$	& 1.075	&	2.42$\times 10^	5	$ &	0                     &	1.13$\times 10^	5	$ &	6.38$\times 10^	4	$   \\ \hline
1.104	&	1.35$\times 10^	5	$ &	8.92$\times 10^	4   $ &	-4.46$\times 10^4	$ &	7.17$\times 10^	4$	& 1.104	&	7.41$\times 10^	4	$ &	0                     &	-1.92$\times 10^5	$ &	6.96$\times 10^	4	$	\\ \hline
1.133	&	3.16$\times 10^	5	$ & 4.87$\times 10^	3	$ &	1.60$\times 10^	5	$ &	7.76$\times 10^	4$	& 1.133	&	8.57$\times 10^	4	$ &	0                     &	-1.86$\times 10^5	$ &	7.53$\times 10^	4	$	\\ \hline
1.162	&	6.94$\times 10^	5	$ &	3.59$\times 10^	5	$ &	5.27$\times 10^	5	$ &	8.37$\times 10^	4$	& 1.162	&	1.13$\times 10^	5	$ &	0	                  &	-1.36$\times 10^5	$ &	8.12$\times 10^	4	$	\\ \hline
1.191	&	1.13$\times 10^	6	$ &	7.65$\times 10^	5	$ &	9.46$\times 10^	5	$ &	9.02$\times 10^	4$	& 1.191	&	4.65$\times 10^	5	$ &	1.15$\times 10^	5	$ &	2.90$\times 10^5	$ &	8.75$\times 10^	4	$	\\ \hline
       
1.365	&	2.63$\times 10^	6	$ &	2.08$\times 10^	6	$ &	2.36$\times 10^	6	$ &	1.36$\times 10^	5$	& 1.365	&	2.32$\times 10^	6	$ &	1.80$\times 10^	6	$ &	2.06$\times 10^	6	$ &	1.32$\times 10^	5	$	\\ \hline
1.394	&	1.50$\times 10^	6	$ &	9.16$\times 10^	5	$ &	1.21$\times 10^6	$ &	1.45$\times 10^	5$	& 1.394	&	7.22$\times 10^	5	$ &	1.60$\times 10^	5	$ &	4.41$\times 10^5   $ &	1.40$\times 10^	5	$	\\ \hline
1.423	&	7.45$\times 10^	5	$ &	1.29$\times 10^	5	$ &	4.37$\times 10^	5	$ &	1.54$\times 10^	5$	& 1.423	&	5.48$\times 10^	5	$ &	0                     &	2.46$\times 10^	5	$ &	1.49$\times 10^	5	$	\\ \hline
1.452	&	2.20$\times 10^	5	$ &	0                     &	-2.95$\times 10^5	$ &	1.63$\times 10^	5$	& 1.452	&	6.58$\times 10^	5	$ & 2.24$\times 10^	4	$ &	3.40$\times 10^	5	$ &	1.59$\times 10^	5	$	\\ \hline
1.481	&	2.18$\times 10^	5	$ &	0                     &	-3.40$\times 10^5	$ &	1.70$\times 10^	5$	& 1.481	&	4.67$\times 10^	5	$ &	0                     &	1.19$\times 10^	5	$ &	1.65$\times 10^	5	$	\\ \hline
1.510	&	4.90$\times 10^	5	$ &	0                     &	1.10$\times 10^5	$ &	1.78$\times 10^	5$	& 1.510	&	6.16$\times 10^	5	$ &	0                     &	2.64$\times 10^	5	$ &	1.73$\times 10^	5	$	\\ \hline
1.539	&	1.02$\times 10^	6	$ & 2.46$\times 10^	5$    &	6.35$\times 10^5	$ &	1.94$\times 10^	5$	& 1.539	&	1.48$\times 10^	6	$ &	7.24$\times 10^	5	$ &	1.10$\times 10^	6	$ &	1.89$\times 10^	5	$	\\ \hline
1.568	&	1.54$\times 10^	6	$ &	7.12$\times 10^	5$    &	1.12$\times 10^	6	$ &	2.06$\times 10^	5$	& 1.568	&	2.05$\times 10^	6	$ &	1.25$\times 10^	6	$ &	1.65$\times 10^	6	$ &	2.00$\times 10^	5	$	\\ \hline
1.597	&	2.32$\times 10^	6	$ &	1.45$\times 10^	6	$ &	1.88$\times 10^	6	$ &	2.18$\times 10^	5$	& 1.597	&	1.73$\times 10^	6	$ &	8.90$\times 10^	5	$ &	1.31$\times 10^	6	$ &	2.11$\times 10^	5	$	\\ \hline

\end{tabular}
\caption{The calculated EoR upper limits and related values for each polarization. We include the $k$ (\textit{h}\,Mpc$^{-1}$), the upper limit $\Delta_U^2$ (mK$^2$), the lower uncertainty bound $\Delta_L^2$ (mK$^2$), the measured power $\Delta^2$ (mK$^2$), and the 1$\sigma$ thermal noise (mK$^2$).}
\label{table:all_eor}

\end{table}

\bibliographystyle{aasjournal}
\bibliography{bib}



\end{document}